\documentclass[12pt, journal, onecolumn]{IEEEtran}
\usepackage[hidelinks]{hyperref}
\usepackage{amsmath, epsfig, cite}
\usepackage{amsthm}
\usepackage{amsfonts}
\usepackage{graphicx}
\usepackage{soul}
\usepackage{latexsym}
\usepackage{amssymb}
\usepackage{color}
\usepackage{url}
\usepackage{colortbl}
\usepackage{comment}
\usepackage[dvipsnames]{xcolor}
\usepackage{caption}
\usepackage{subcaption}

\usepackage{cite}
\usepackage{hyperref}
\usepackage{cleveref}

\usepackage{xfrac}
\usepackage{diagbox}

\DeclareMathAlphabet{\mathbfsl}{OT1}{ppl}{b}{it} 

\newcommand{\C}{\mathbb{C}}

\newcommand{\N}{\mathbb{N}}

\newcommand{\R}{\mathbb{R}}

\newcommand{\E}{\mathbb{E}}

\newcommand{\cC}{{\cal C}}
\newcommand{\cD}{{\cal D}}
\newcommand{\cE}{{\cal E}}

\newcommand{\cL}{{\cal L}}

\newcommand{\cO}{{\cal O}}

\newcommand{\cS}{{\cal S}}

\newcommand{\cX}{{\cal X}}
\newcommand{\cY}{{\cal Y}}

\newcommand{\bfp}{{\boldsymbol p}}

\newcommand{\bfu}{{\boldsymbol u}}
\newcommand{\bfv}{{\boldsymbol v}}

\newcommand{\bfx}{{\boldsymbol x}}
\newcommand{\bfy}{{\boldsymbol y}}

\newcommand{\bfU}{{\mathbf U}}
\newcommand{\bfV}{{\mathbf V}}

\newcommand{\bfX}{{\mathbf X}}

\theoremstyle{definition}
\newtheorem{theorem}{Theorem}
\newtheorem{lemma}{Lemma}
\newtheorem{remark}{Remark}
\newtheorem{construction}{Construction}
\newtheorem{corollary}{Corollary}
\newtheorem{definition}{Definition}
\newtheorem{conjecture}{Conjecture}
\newtheorem{example}{Example}

\newtheorem{problem}{Problem}
\newtheorem{claim}{Claim}

\newcommand{\abs}[1]{|#1|}

\begin{document}

\title{\textbf{Cover Your Bases: How to Minimize the Sequencing Coverage in DNA Storage Systems}}

\author{\textbf{Daniella~Bar-Lev},~\IEEEmembership{Student~Member,~IEEE,}
        \textbf{Omer~Sabary},~\IEEEmembership{Student~Member,~IEEE,}
        \textbf{Ryan~Gabrys},~\IEEEmembership{Senior~Member,~IEEE,}
        and\textbf{~Eitan~Yaakobi},~\IEEEmembership{Senior~Member,~IEEE}
  \thanks{Parts of this work were presented at the IEEE International Symposium on Information Theory (ISIT), Taipei, Taiwan, 2023~\cite{BS23}.}
  \thanks{D. Bar-Lev, O. Sabary and E. Yaakobi are with the Henry and Marilyn Taub Faculty of Computer Science, Technion - Israel Institute of Technology, Haifa 3200003, Israel 
  (e-mail: \texttt{\{daniellalev,omersabary,\linebreak yaakobi\}@cs.technion.ac.il}). 
  R. Gabrys is with University of California, San Diego, California, USA   
  (e-mail: \texttt{rgabrys@ucsd.edu}).}
  \thanks{%
  The research was funded by the European Union (ERC, DNAStorage, 865630). Views and opinions expressed are however those of the authors only and do not necessarily reflect those of the European Union or the European Research Council Executive Agency. Neither the European Union nor the granting authority can be held responsible for them. This work was also supported in part by NSF Grant CCF2212437.}
  \thanks{%
  The first two authors contributed equally to this work.}}

\maketitle
\begin{abstract}
Although the expenses associated with  DNA sequencing have been rapidly decreasing, the current cost of sequencing information 
stands at roughly \$120/GB, which is dramatically more expensive than reading from existing archival storage solutions today. 
In this work, we aim to reduce not only the cost but also the latency of DNA storage by initiating the study of the \textit{\textbf{DNA coverage depth problem}}, which aims to reduce the required number of reads to retrieve information from the storage system. Under this framework, our main goal is to understand the effect of error-correcting codes and retrieval algorithms
on the required sequencing coverage depth.  We establish that the expected number of reads that are required for information retrieval is minimized when the channel follows a uniform distribution. We also derive upper and lower bounds on the probability distribution of this number of required reads and provide a comprehensive upper and lower bound on its expected value. We further prove that for a noiseless channel and uniform distribution, MDS codes are optimal in terms of minimizing the expected number of reads.
Additionally, we study the DNA coverage depth problem under the random-access setup, in which the user aims to retrieve just a specific information unit from the entire DNA storage system.  
We prove that the expected retrieval time is at least $k$ for $[n,k]$ MDS codes as well as for other families of codes. Furthermore, we present explicit code constructions that achieve expected retrieval times below $k$ and evaluate their performance through analytical methods and simulations. Lastly, we provide lower bounds on the maximum expected retrieval time. Our findings offer valuable insights for reducing the cost and latency of DNA storage.
\end{abstract}

\section{Introduction} \label{sec:intro}

The world's digital data is growing exponentially, doubling from 30 to 64 zettabytes in just three years, and it is anticipated to reach 180 zettabytes by 2025, resulting in a data storage crisis. The demand for storage capacity already exceeds the supply, and the gap continues to grow~\cite{IDC}. Recent research and insights from the IDC emphasize the struggle of existing storage technologies to meet the demands of the big data era.

Recognizing this challenge, DNA emerges as a promising storage medium due to its exceptional density and durability. 
The DNA storage pipeline usually involves three main components. The first is \emph{DNA synthesis}, which produces artificial DNA molecules. These synthetic DNA molecules are called \emph{oligos} or \emph{strands} and they can be designed in a way that encodes the user's information. The current synthesis technologies only produce strands that are up to a length of 300 bases~\cite{LP10} and due to technology limitations, they also produce several noisy copies per encoded strand. Thus, it is likely that the user information is stored in several different strands. The second component of the DNA storage pipeline is a \emph{storage container}, usually a small tube that contains all the short strands that encode the user information. Lastly, to read back the user information, it is required to perform \emph{DNA sequencing} on the strands in the tube. The sequencing process translates the DNA strands into digital sequences over the DNA alphabet, which are noisy copies of the synthesized strands. These DNA sequences can be decoded to read back the user's information. 

The sequencing process, which is done using a DNA sequencer, is one of the principal components in any DNA storage system~\cite{EZ17, OAC17, YGM17, AVA19}. Nowadays, DNA sequencers suffer from relatively slow throughput as well as high costs relative to other alternative storage technologies~\cite{Whitepaper, SH22, YK15}. These issues are related to the so-called \emph{coverage depth} of DNA storage, which is defined as the ratio between the number of reads that are sequenced and the number of designed strands~\cite{HMG19}. Reducing the coverage depth can improve the latency of any existing DNA storage system and reduce its costs. 

Motivated by the connection between the coverage depth, latency, and cost,  and in an effort to design coding schemes that overcome the drawbacks associated with existing sequencing technologies, in this work we initiate the study of a novel problem, referred to as the  \emph{DNA coverage depth problem}.
Simply stated, the DNA coverage depth problem 
 aims to minimize the coverage depth while maintaining system reliability. We will study the required coverage depth as a function of the DNA storage channel, the error-correcting code, and the algorithms involved in retrieving the user's information.
 Furthermore, we seek to understand how to pair an error-correcting code with a given DNA storage system in order to minimize the coverage depth. 
This problem will be studied under both the random and non-random access settings. While the latter addresses the problem of retrieving all the information that was being stored, the former describes the case in which the user is interested in retrieving only a specific part of the stored information. Moreover, we plan to suggest coding schemes that optimize the required coverage depth and to study, both theoretically and experimentally, how one can utilize codes to minimize the sequencing time and costs. 

Despite significant work on DNA storage, only a small number of works have focused on reducing the latency and costs associated with sequencing in experimental or theoretical setups. Erlich et. al.~\cite{EZ17} encoded digital information into DNA strands using a Luby transform-based coding scheme. Later, they diluted their synthesized strands and studied the effect of this dilution on their ability to sequence and decode the information. The dilution procedure reduced the potential (maximal) coverage depth of their system down to roughly 1300 reads per strand, thus making the decoding process more challenging. They showed that thanks to the error-correcting capability of their scheme, they were able to perfectly retrieve the stored information. In another related work, Chandak et. al.~\cite{CT19} defined the ratio between the number of synthesized bits and the number of information bits as the \emph{writing cost}, and similarly the ratio between the number of bits that have to be read (sequenced) and the number of information bits was defined as the \emph{reading cost}. In their work, they studied the tradeoffs and relations between the writing and reading costs. They first showed that for the noiseless channel, it is enough to read one copy per designed strand. Thus, the relation of these two costs can be obtained by inferring the channel as an erasure channel with an erasure probability that can be approximated using Poisson approximation. Additionally, the authors suggested an LDPC-based coding scheme that can improve the ratio between the two costs. They also showed by simulations how their suggested scheme can be used with different redundancy levels to reduce both the writing cost and the reading cost.

The DNA coverage depth problem is related to 
the \emph{coupon collector's (CCP)},  
\emph{dixie cup}, and \emph{urn} problems~\cite{ErdosReyni,FGT92, NewmanDoubleDix,Feller}. For all these problems, it is assumed that there are~$n$ different types of coupons and the question of interest is \emph{how many coupons one should collect before possessing one coupon of each type}. It is well known that if the coupons are drawn uniformly at random (with repetition), then the expected number of coupons necessary 
to have at least one coupon from each type is roughly $n\log n$. Under our setting, the coupons refer to the copies of the synthesized oligos and the goal is to read at least one copy of every oligo.

The CCP has several generalizations~\cite{ErdosReyni, FGT92, NewmanDoubleDix}, some of which will be explored in this work. One such problem, which is referred to as the \emph{MDS coverage depth} problem, is
\emph{how many coupons one should collect before possessing $t$ copies of $k$ coupons}. This generalization represents the scenario where a reconstruction algorithm that requires $t$ reads of an oligo for successful decoding is used along with an MDS code that requires correctly retrieving $k$ out of the $n$ synthesized sequences to recover the stored encoded information. Another problem that is addressed in this paper is the \emph{coding coverage depth problem}, which generalizes the MDS coverage depth and considers the effect of an error-correcting code, which is not necessarily an MDS code. Under this setup, our main results show that MDS codes are optimal codes for the purpose of reducing the expected coverage depth. Furthermore, our analysis for the MDS coverage depth problem provides a deep understanding of the required number of reads that should be sequenced in order to guarantee a successful retrieval of the information with high probability.    

Additionally, motivated by the random-access setting where one wishes to retrieve a single strand of DNA from a storage system, in \autoref{sec:random} we consider another problem that is related to the CCP, but to the best of our knowledge has not been studied before. Suppose we are given $k$ information coupons which we can encode into a set of $n$ total coupons.  \emph{For any information coupon say $i$, what is the expected number of coupons that need to be collected in order to retrieve the information in coupon $i$?} Trivially, if no code is used and every coupon is collected with the same probability, then the expected number of coupons that need to be collected is equal to $k$. In \autoref{sec:random}, we initiate the study of this problem, which we refer to as the \emph{singleton-random-access problem}. Our main result is to show that it is indeed possible to design coding schemes that allow random access that requires less than $k$ coupons and provide examples of several such schemes.

This paper is organized as follows.  \autoref{sec:defs} introduces the definitions that are used throughout the paper. In \autoref{sec:sequencing coverage}, we formally define the problems that are studied throughout this paper along with related work. \autoref{sec:sequencing coverage} also gives a detailed summary of the main results given in this paper. Next, in \autoref{sec:noiseless}, we consider the case in which the channel is noiseless and address the MDS coverage depth problem and the coding coverage depth problem for the noiseless channel.  \autoref{sec:sequencing coverage:erroneous} extends the study of the MDS coverage depth problem to noisy channels, and gives several bounds on the success probability of the decoding as a function of the number of reads that were sequenced. Finally, in  \autoref{sec:random}, we present our results for the singleton-random-access problem. For more details on the results and contributions presented in each section, the reader is referred to \autoref{subsec:main_contribution}.

\section{Definitions and Channel Model}
\label{sec:defs}
\begin{figure}
\includegraphics[width=\linewidth]{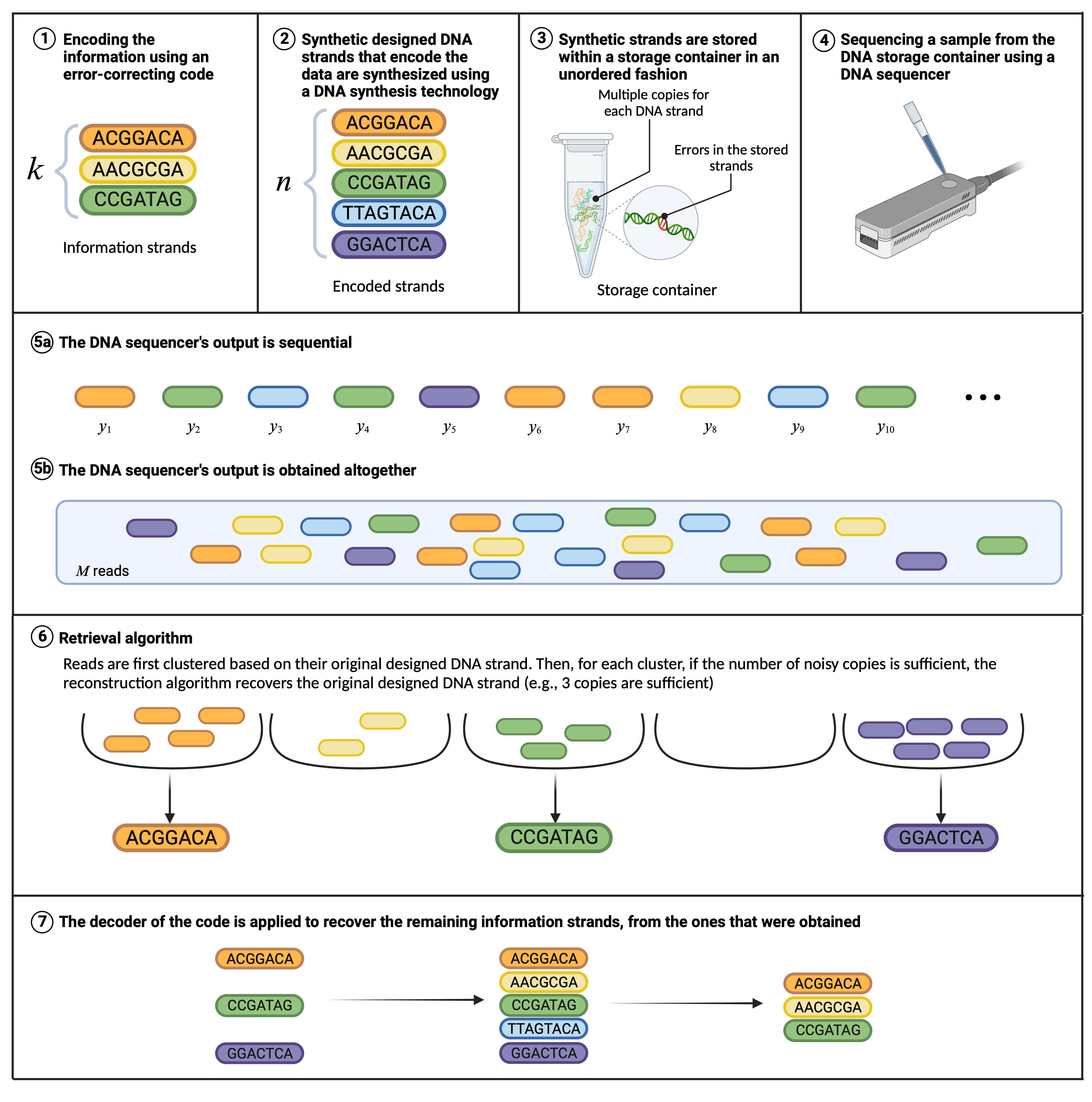}
\caption{{The DNA storage pipeline.} }\label{fig:system_description}
\end{figure}

In the typical model of DNA-based storage systems~\cite{YGM17, OAC17, EZ17}, the data is stored as a codeword that can be described by a vector of length-$\ell$ \emph{sequences} or \emph{strands} over the alphabet $\Sigma=\{A, C, G, T\}$. The set of all length-$\ell$ vectors over $\Sigma$ is denoted by $\Sigma^\ell$, and $\Sigma^* \triangleq \bigcup_{\ell=0}^\infty \Sigma^\ell$. For a positive integer $n$, $[n]$ denotes the set $\{1, \ldots, n\}.$ In many cases an \emph{outer error-correcting $\cC$} is used to encode the data over the {length-$\ell$} sequences, so it is assumed that the outer code $\cC$ receives a vector of $k$ {length-$\ell$} sequences, $\bfU = ( \bfu_1, \bfu_2, \ldots, \bfu_k )\in (\Sigma^\ell)^k$ and returns a vector of $n$ length-$\ell$ sequences $\bfX = ( \bfx_1, \bfx_2, \ldots, \bfx_n )\in (\Sigma^\ell)^n$. For two vectors $\bfU = (\bfu_1, \ldots, \bfu_{k_1} )$ and $\bfV = (\bfv_1, \ldots, \bfv_{k_2})$, we denote by $\bfU \circ \bfV$ their \emph{concatenated vector}, i.e., $\bfU \circ \bfV = (\bfu_1, \ldots, \bfu_{k_1}, \bfv_1, \ldots, \bfv_{k_2})$. In this work, the code $\cC$ is denoted by $(n,k)$ or by $[n,k]$ in case $\cC$ is an MDS code.
The vector $\bfX$ is the input to the DNA storage system, which we now describe in more detail and is also illustrated in \autoref{fig:system_description}. 

The DNA storage channel, denoted by $\mathsf{S}$, first produces many noisy copies for each of the strands in the vector $\bfX$. Then, these noisy copies are amplified using PCR, and lastly, a \emph{sample} of $M$ of these strands is sequenced using a DNA sequencing technology\cite{HMG19}. Therefore, the output of the DNA storage channel can be described as a multiset $\cY_M =\{\!\!\{ \bfy_1, \bfy_2, \ldots, \bfy_M \}\!\!\}$, where each $\bfy_j\in \Sigma^*$ for $j \in [M]$ is called a \emph{read} and is a noisy version of some $\bfx_i$, $i\in[n]$. It should be noted that our model assumes that for any read $\bfy_j$, the index $i\in[n]$ such that $\bfy_j$ is a noisy copy of $\bfx_i$ is known (this can be achieved by encoding the index $i$ within the strand $\bfx+i$). Depending on the specific sequencing technology being used, the reads in $\cY_M$ can be obtained either sequentially (one after the other), or altogether. The former corresponds to Nanopore sequencing technologies~\cite{WZ21}, while the latter describes next-generation sequencing (NGS) technologies~\cite{BD14} (e.g. Illumina). The number of reads in $\cY_M$ that are noisy copies of the $i$-th strand $\bfx_i, i\in [n]$, depends upon some categorical probability distribution $\bfp=(p_1,\ldots,p_n)$, where for $i\in[n]$, $p_i$ is the probability to sample a read of $\bfx_i$. The probability distribution $\bfp$ is a function of the DNA storage channel $\mathsf{S}$ and is referred by the \emph{channel probability distribution}, or in short \emph{channel distribution}; Note that the distribution $\bfp$ might also depend on the design of the DNA strands in $\bfX$, however for simplicity, in this work we assume that $\bfp$ is only a function of the channel $\mathsf{S}$. 

\begin{remark} \label{remark_1}
Note that in several works, see e.g.~\cite{SR19, LS20}, it is assumed that a \emph{set} (and not a vector) of strands is stored in the DNA storage system. However, since the strands in these sets are  tagged by indices anyway, we assume for simplicity that the information is a vector of strands. Furthermore, it may also be possible that every strand is encoded using an inner code~\cite{EZ17, OAC17}. Nevertheless, since this part is independent of the study of this work, it is not treated as part of the encoding process, but it is taken into account in the success probability of a retrieval algorithm, as will be explained below. 
\end{remark}

The decoding process of $\bfX$ (and thus $\bfU$) starts with partitioning the reads in $\cY_M$ into groups, also called \emph{clusters}, according to their origin strand, i.e., for $i\in [n]$, the $i$-th cluster should contain all the reads $\bfy_j$ that are noisy copies of  $\bfx_i$. To simplify the analysis, we assume that this step is accomplished error-free. In practice, this assumption can be reached using indices in the sequence $\bfx_i$ which can be further protected using some error-correcting code\cite{YGM17}. Hence, the probability of successfully retrieving $\bfX$ and $\bfU$ mainly depends on the following two components of the solution being used. 

\begin{enumerate}
\item \emph{Error-correcting code}.  When $\bfX$ is a codeword in some error-correcting code $\cC$, it is possible to successfully retrieve $\bfX$ even if not all of its $n$ symbols were decoded successfully. The applicable subsets $J \subseteq [n]$ such that $\bfX$ can be retrieved from the symbols $\bfx_j$ for $j\in J$ are determined by the code $\cC$. For example, if $\cC$ is an $[n,k]$ MDS code, then any $k$ strands (symbols of $\bfX$) are sufficient to decode the data.

\item \emph{The retrieval algorithm}. The success probability to retrieve the strand $\bfx_i$ also depends on the \emph{retrieval algorithm}, which aims to decode a sequence using several noisy copies~\cite{BP21}. Typically, this probability depends on the number of noisy copies which are given as input, the channel error rates, and the use of an inner code within the strands. 
In this work, we model the retrieval algorithm using an integer $t \ge 1$, and we assume that each strand $\bfx_i$ can be retrieved given $t$ reads, which are noisy copies of it, and cannot be retrieved given less than $t$ reads\footnote{Note that, in practice, the probability that the retrieval algorithm succeeds is not binary. More precisely it is a function that returns a value between $0$ and $1$ and increases with $t$.}.  

\end{enumerate}
The main goal of this paper is to study the required sample size $M$ that guarantees successful decoding of the information. According to our model, this sample size depends on the channel, the error-correcting code, and the channel probability distribution $\bfp$.
\begin{remark} 
The analysis presented in this work assumes that the reads in the multiset $\cY_M$ are received sequentially from the DNA storage channel as illustrated in step 5a of \autoref{fig:system_description}. However, our results are also relevant for the case in which all the reads are obtained together. More specifically, the random variable that governs the sample size $M$ for which decoding is possible in the sequential case can be used to describe the non-sequential case as well. That is, the probability distribution of the latter corresponds to the decoding success probability given $M$ strands in the non-sequential case. 
\end{remark}


In this paper, we explore two different scenarios concerning our problem. In the first scenario, discussed in both Section~\ref{sec:noiseless} and Section~\ref{sec:sequencing coverage:erroneous}, we focus on the objective of recovering all the stored information. This involves retrieving the entire vector $\bfU$. On the other hand, in Section~\ref{sec:random}, we shift our attention to a different scenario where our goal is to retrieve a specific part of the information, i.e., a specific subset of symbols from the vector $\bfU$. For these scenarios, we calculate the expected required sample size for noiseless/noisy channels and study how it can be minimized using 
coding schemes. 

\section{The Coverage Depth Problem in the DNA Storage Channel}\label{sec:sequencing coverage}

\subsection{Problems Definition}

This work studies the required sample size to retrieve the information vector $\bfU$, or a specific subset of its symbols, as a function of the DNA storage channel, the error-correcting code, and the retrieval algorithm. Under this framework, our goal is to understand how to optimally pair an error-correcting code with a given retrieval algorithm in order to minimize the sample size, while guaranteeing successful decoding with high probability. 

 According to our model characterization, we let $\nu^{\bfp}_{t}(\cC)$ be the random variable that governs the number of reads that should be sampled for successful decoding of $\bfU$. When $\cC$ is an $[n,k]$ MDS code, this notation is replaced by $\nu^{\bfp}_{t}(n,k)$. The uniform distribution is denoted by $\bfp_u\triangleq(\frac{1}{n}, \ldots, \frac{1}{n})$ and for brevity, we let $\nu_{t}(\cC) \triangleq \nu^{\bfp_u}_{t}(\cC)$ and $\nu_{t}(n,k) \triangleq \nu^{\bfp_u}_{t}(n,k)$. The first two problems, which focus on retrieving the entire information vector $\bfU$, are defined below.

\begin{problem}(\textbf{The MDS coverage depth problem.})  \label{pr:expectation_prob} 
 For given values of $k$ and $n$, and a channel distribution $\bfp$ find the expectation and the probability distribution of the random variable $\nu_{t}^{\bfp}(n,k)$. That is, find the values  of $\E \left[ \nu_{t}^{\bfp}(n,k) \right]$ and ${P [ \nu_{t}^{\bfp}(n,k) > m]}$ for any $m \in \N$.
\end{problem}

\begin{problem} \label{pr:opt_code}(\textbf{The coding coverage depth problem.}) 
For a given value of $k$, find the following. 
\begin{enumerate} 
    \item Given $n$ and $\bfp$, find an $(n,k)$ code $\cC$ that is optimal with respect to minimizing $\E \left[ \nu^{\bfp}_{t}(\cC) \right]$. 
    \item The minimum value of $\E \left[ \nu^{\bfp}_{t}(\cC) \right]$ over all possible codes $\cC$ with dimension $k$ and channel distributions $\bfp$.
That is, find the value
$ \mathsf{M}^{opt} (k) \triangleq \liminf_{\cC, \bfp} \{\E \left[ \nu^{\bfp}_{t}(\cC) \right]\}.$
\end{enumerate}
  
\end{problem}

The third problem is related to the other setup, in which the user wishes to retrieve a subset of the $k$ information strands (i.e., a subset of $\bfU$'s symbols).
This subset can be described by an index set $I \subseteq [k]$, such that the set of information strands to be retrieved is $\bfU_I =\{\bfu_i : \  i \in I \}$. 
In this work, we consider the special case in which this subset is a singleton, i.e., the case where the user wishes to retrieve a single information strand $\bfu_i$ for some $i \in [k]$. 
More formally, we are interested in the following problem. 

\begin{problem}\label{prob:random:single}(\textbf{The singleton coverage depth problem.}) 

Given an $(n,k)$ code $\cC$, for ${i\in [k]}$, let $\tau_i(\cC)$ be the random variable that governs the number of samples to recover the $i$-th information strand assuming noiseless channel with uniform distribution. Find the following:
\begin{enumerate}
\item The expectation value $\E [ \tau_i(\cC)] $ and the probability distribution $P[\tau_i(\cC)>r]$ for any $r\in\mathbb{N}$. 
\item The maximal expected number of samples to retrieve an information strand, i.e., $$T_{\max}^\cC \triangleq \max_{1\le i\le k} \mathbb{E}[\tau_i(\cC)].$$
\item The average expected number of trials to retrieve an information strand, i.e., $${T_{\textrm{avg}}^\cC \triangleq  \frac{1}{k}\sum_ {i=1}^{k} \mathbb{E}[\tau_i(\cC)]}.$$
\end{enumerate} 
When no coding is used, (i.e., $\bfU=\bfX$) 
$\cC$ is removed from the notations.
\end{problem}

\subsection{Related Work}
\label{sec:sequencing coverage:noiseless} 
For the noiseless channel, it is sufficient to have a single read of each $\bfx_i,i\in[n]$ to retrieve it.
We note that if the channel distribution is the uniform distribution $\bfp_u$, and no code is defined on the data (i.e., $k=n$) then finding the expectation listed in Problem~\ref{pr:expectation_prob} is equivalent to the classical \emph{coupon collector's problem}~\cite{Feller}. This problem was first studied by Feller~\cite{Feller} where it was referred to as the \emph{dixie cup problem}. Under the assumption that we have $n$ coupons and it is equally likely to collect any of the coupons, the expected number of draws (i.e., sample size) required to get a single copy for each coupon is $\E[ \nu_1 (n,k=n)] = nH_n =  n \log n + \gamma n + \cO(1)$,  where $H_n$ is the $n$-th harmonic number and $\gamma \sim 0.577$ is the Euler–Mascheroni constant. Furthermore, it was also proven~\cite{FGT92} that
$\E [\nu_1(n,k)]  = n(H_n-H_{n-k}) $. It is well-known that when $\lim_{n \to \infty} n-k  = \infty$,\footnote{In this case, there exists $0<a<1$, such that for $n$ large enough $k < an$. } the expectation can be approximated by $\E [\nu_1(n,k)] \approx n\log(n) - n \log(n-k) = n \log(\frac{n}{n-k}).$

For noisy channels, i.e., $t>1$, the problem is closely related to the classical \emph{urn problem}~\cite{ErdosReyni, NewmanDoubleDix}. Suppose there are $n$ labeled urns and each can be filled with identical balls. At every round, a ball is thrown into one of the urns randomly. In each round, the probability of throwing a ball to the $j$-th urn is denoted by $p_j$, for $1\le j \le n$, and we let $\bfp=(p_1, \ldots, p_n)$. In~\cite{NewmanDoubleDix}, it was shown that in order to have $t$ balls in each urn (or equivalently $t$ copies per coupon), the expected sample size is  $\E[ \nu_t (n,k=n)] = n \log n + n(t-1) \log \log n + n C_t + o(n)$, where $C_t$ is a constant that depends on $t$. Following that, Erd\H{o}s and R\'{e}nyi~\cite{ErdosReyni} proved that the distribution of this random variable is tightly concentrated around the expectation. More specifically, when drawing $n\log n + n(t-1)\log \log n +nx$ times, the probability to have at least $t$ copies for $n$ coupons is asymptotically equal to  $e^{-\frac{e^{-x}}{(t-1)!}}$ for $n$ large enough. 
Flajolet et al.~\cite{FGT92} generalized these results to a general discrete distribution on the coupons/balls and proved that the expected sample size to have at least $t$ copies/balls for $k$ out of the $n$ coupons/urns is

\vspace{-2.5ex} 
\begin{small}
\begin{align} \label{eq:gen_t_k}
\E[ \nu_{t}^{\bfp} (n,k)]  \hspace{-0.77ex}=\hspace{-0.85ex} \sum_{q=0}^{k-1}\hspace{-0.5ex} \int_0^\infty\hspace{-1.5ex} [u^q] \prod_{i=1}^n \hspace{-0.4ex}\left( e_{t-1} (p_i v) \hspace{-0.5ex}+\hspace{-0.5ex}u\left(e^{p_i v} \hspace{-0.5ex}-\hspace{-0.5ex}e_{t-1} (p_i v)  \right) \right) \hspace{-0.4ex}e^{-v}\hspace{-0.35ex} dv,  \vspace{-1.05ex} \end{align}
\end{small}
where $e_t(x)=\sum_{i=0}^t \frac{x^i}{i!}$ 
and for a polynomial $Q(u)$, $[u^q]Q(u)$ is the coefficient of $u^q$ in $Q(u)$. This known result solves the expectation value listed in \autoref{pr:expectation_prob}, not only for $\bfp_u$ but for any channel distribution.  
As can be seen, for practical purposes, the expression in (\ref{eq:gen_t_k}) and its asymptotic behavior are not easy to calculate or to work with. Hence, in \autoref{sec:sequencing coverage:erroneous} we solve a closely related problem and present a closed-form expression. Moreover, to the best of our knowledge, the other part of \autoref{pr:expectation_prob}, i.e., studying the cumulative probability distribution ${P [ \nu_{t}^{\bfp}(n,k) > m]}$, is still open. 

Another related problem was presented in~\cite{CT19} by Chandak et al. In their paper, the authors defined the \emph{writing cost} as the number of synthesized bases per information bit, and the \emph{reading cost} as the number of bases that have to be sequenced per information bit in order to retrieve the stored information. Their paper studies the tradeoffs between these two costs. They first showed that for the noiseless channels, the event of obtaining zero copies of a specific strand is equivalent to an erasure of this strand, which can be approximated as a Poisson random variable. Thus, they were able to compute the capacity of this channel and by this obtaining the tradeoffs of the costs. For the noisy channel, the authors suggested an LDPC-based scheme to improve the ratio between the costs, for more details see~\cite{CT19}. 

The problem of random access in DNA storage has already been addressed in several works; see e.g.~\cite{BSB_etal21,LCR_etal21,OAC17,YYM_etal15,WOW_etal22}. The main goal is to support random access to specific DNA strands in the storage and this can be supported by the use of different primers for the different strands or physically storing strands in different storage containers. However, these solutions incur high costs, and thus the problem of storing strands together using these primers is still important and this work addresses it from a coding theory perspective.


\subsection{Main Contributions} \label{subsec:main_contribution}

In this paper, we define a new family of problems that should be considered when designing DNA storage systems. Additionally, this work provides an extensive analysis and present results that enhance our understanding of the interplay between error-correcting codes, retrieval algorithms, and coverage depth. The main results with respect to each of the three problems we defined are listed below.

\noindent \textbf{The MDS coverage depth problem (\autoref{pr:expectation_prob})} For this problem, we have the following results. 
\begin{enumerate}
    \item 
    We show in~\autoref{lem:uni-opt} that the value of $\E [\nu_t^\bfp(n,k)]$ is minimized if and only if the channel has the uniform distribution. 

    \item We show in~\autoref{th:concentrate} and in~\autoref{th:concentrate_fixed} two upper bounds on the probability distribution of $P[\nu_t(n,k) > m]$.
    We further prove in~\autoref{cl:lower_bound_prob} a lower bound on the probability $P[\nu_t(n,k) \le~m]$. 
    Combining these results in~\autoref{th:lower_upper_exp} we prove that for any $\varepsilon>0$,
    
    \vspace{-1ex}
    \begin{small}
    $$  \log\left(\frac{1}{1-R} \right) +f_c(n,R)   \le \E \left[ \frac{\nu_t (n,k)}{n} \right] \le  \left(  \log \left(\frac{1}{1-R}\right) +  t \log \log n + 2  \log (t+1) \right)\cdot (1+ 2\varepsilon ) , $$   
    \end{small}
    where $f_c(n,R) = \cO(\frac{1}{n^2}).$

    \item For practical purposes of DNA storage systems, it is sometimes required to plan ahead and set the number of reads that should be sampled to guarantee successful decoding. Hence, we show in~\autoref{the: lower bound r}, that when sampling more than $r_E(n,k,t)$ reads, the expected number of encoded strands that can not be recovered (i.e., have less than $t$ copies) is at most $n-k$, which indicates on the probability of successful decoding. The value of $r_E(n,k,t)$ can be found in equation (\ref{eq:r_E}).
\end{enumerate}

\noindent \textbf{The coding coverage depth problem (\autoref{pr:opt_code})}
We fully solve \autoref{pr:opt_code} for the noiseless channel with uniform distribution. We show that MDS codes are optimal with respect to minimizing $\E[\nu_t(n,k)]$. We also show that for a fixed $k$, the larger $n$ is, the smaller the value of $\E[\nu_t(n,k)]$ is. The results of this problem are given in~\autoref{th:MDS} and~\autoref{th:optimal_ex}.

\noindent \textbf{The singleton coverage depth problem (\autoref{prob:random:single})}
We extensively study the singleton coverage depth problem for the case in which the channel is noiseless. Our main results are summarized below.   
\begin{enumerate}
    \item We first study and fully solve the case in which $n=k$. In particular, we prove that if $n=k$ then the expected time to retrieve a singleton is minimized when no coding is used and it is equal to $k$ and $T_{\max}^{\cC} = T_{\text{avg}}^{\cC}=  k$ (see \autoref{lem:random:single:no code} and \autoref{cl:ra:n=k_nocode}).
    \item Next, to study more involved cases, we define \emph{retrieval sets} and \emph{minimal retrieval sets}, which correspond to the (minimal) sets of encoded strands from which a specific target singleton information strand can be recovered. Using this property of codes, we analyze the expected time to retrieve a singleton given that its minimal retrieval sets are disjoint. See these results in \autoref{theorem:random:single:two disjoint retrieval sets} and \autoref{corollary:random:single: disjoint retrieval sets}. Moreover, in \autoref{corollary:simple-parity} we use  \autoref{theorem:random:single:two disjoint retrieval sets} to conclude that the expected time to retrieve a singleton given that $\cC$ is the simple parity $[k+1,k]$ code is $k$.
    \item We extend the result of \autoref{corollary:simple-parity} to any systematic $[n,k]$ MDS code, by the construction and detailed evaluation of the corresponding generating function. That is, we show in \autoref{th:MDS_random_access}, that for any $[n,k]$ MDS code $\cC$ and any $i\in [k]$, $\E [\tau_i(\cC) ] = T_{\max}^{\cC} = T_{\text{avg}}^{\cC}=  k$.
    \item We give two explicit code constructions (\autoref{construction_2k_k} and \autoref{construction_ryan}) for codes with $k$ information strands for which $T_{\max}^{\cC} <k$, i.e., $\E [\tau_i(\cC) ] <k $ for all $i \in [k]$. Furthermore, we analyze the behavior of these codes both analytically and by computer simulations.
    \item To conclude the analysis of the singleton coverage depth problem, we provide in \autoref{lm:lowerboundra} and in \autoref{th: ramdom access: improved lower bound} two lower bounds on the value of $T_{\max}^{\cC}$. Moreover, in \autoref{corollary:lower bound rate}, for $n$ large enough, we show that for any $(n,k)$ code $\cC$, such that $R = \frac{k}{n}$, we have that ${T_{\max}^{\cC}} \ge k\left(\frac{1}{R}+\frac{1-R}{R^2}\cdot \ln(1-R)\right)$. In particular, the latter implies that when $R$ approaches zero, ${T_{\max}^{\cC}} \ge \frac{k}{2}$, and when $R$ approaches one, the lower bound approaches $k$ from below. 
\end{enumerate}

\section{The Coding Coverage Depth Problem - Noiseless Channel} \label{sec:noiseless}
In this section, we focus on the setup where the channel is noiseless which refers to $t=1$.
Hence, we can assume that the retrieval algorithm simply returns the sampled sequences and thus if $\bfx_i$ has at least one copy, i.e., $t \ge 1$, it is enough to retrieve it. Under this setup, the minimum sample size $M$ is equivalent to the quantity which is governed by the random variable $\nu^{\bfp}_{1}(n,k)$. Using our notation, note that the expected value of $\nu^{\bfp}_1(n,k)$ is given in~(\ref{general_p_t_1}). In this case the distribution probability function was studied in~\cite{ErdosReyni} and is given in~(\ref{eq:gen_t_k}). 
Clearly, when $k=1$, we have that $\E[\nu_{1}(n,1)]=1$. Hence, this section is focused on the case where $k \ge 2$. 
Our main result is to solve Problem~\ref{pr:opt_code} and to show that MDS codes are optimal for any categorical channel distribution. Furthermore, we show that $\E [\nu^{\bfp}_{1}(n,k)]$ is minimized when $\bfp$ is the uniform distribution and is bounded from below by $k \log e$ if $\frac{k}{n} = \Theta(1)$.

In light of the existing results and as a first step toward obtaining Theorem~\ref{th:optimal_ex}, we first show that for the uniform channel distribution, when $k$ is fixed, $\E [\nu_1(n,k)]$ decreases as $n$ increases. 
\begin{claim} \label{cl:dec-function}
For all $n \geq k$, 
$\E [\nu_1(n,k)] > \E [\nu_1(n+1,k)].$
  \end{claim}
  \begin{proof}
The proof follows by showing that $\E [\nu_1(n,k)]$ is a monotonic function that decreases with~$n$. 
 From~\cite{FGT92} for any $n \in \N$ we have that 
\begin{align*}
 & \E [\nu_1(n,k)] - \E [\nu_1(n+1,k)]    = \sum_{i=0}^{k-1}\frac{n}{n-i} - \sum_{i=0}^{k-1}\frac{n+1}{n+1-i}\\
 & = \sum_{i=0}^{k-1}\left(\frac{n}{n-i}  - \frac{n+1}{n+1-i} \right) =  \sum_{i=0}^{k-1} \frac{i}{(n-i)(n+1-i)} > 0,
 \end{align*}
 which completes the proof.
  \end{proof}
  
The next claim solves \autoref{pr:opt_code}.1 and states that given $k$ information strands, for any channel distribution $\bfp$, using an $[n,k]$ MDS code minimizes the expectation of $
\nu_1^{
\bfp
}(\cC)$  compared to any other length-$n$ codes. This can be verified by showing that the number of subsets of size $k$ which are sufficient to retrieve the information is maximized when an MDS code is used. 
 \begin{claim} \label{cl:mds-opt}
 Given $k$, $n$, and $\bfp$, assume that $\cC$ is an $(n,k)$ code. Then, it holds that, $\E [\nu_1^{\bfp}(n,k)] \le \E [\nu_1^{\bfp} (\cC)],$ where equality is obtained if and only if $\cC$ is an MDS code. 
 \end{claim}
 \begin{proof}
Given a sample of size $M$, we denote by $J \subseteq [n]$ the indices of the unique strands that are represented in this sample. If $|J| <k$ then it is impossible to successfully decode the information, which follows since the dimension of the code is $k$.  Otherwise, when $|J| \ge k$, any $[n,k]$ MDS code can decode the stored information, while if $\cC$ is not an MDS code there exists $J'$ of size $k$ from which the stored information can not be decoded using $\cC$. 
Therefore, if $\cC$ is not an MDS code, for any $J \subseteq [n]$ we have that either none of the codes can successfully decode the information, or that the $[n,k]$ MDS code can decode, while $\cC$ cannot. This implies the inequality stated in the theorem, where equality holds if and only if $\cC$ is an MDS code. 
 \end{proof}

We continue towards solving \autoref{pr:opt_code}.2, and in the next theorem it is shown that for MDS codes, $\E [\nu^{\bfp}_1(n,k)]$ is minimized when $\bfp = \bfp_u$.

\begin{theorem} \label{lem:uni-opt} 
For any $\bfp$,
$ \E [\nu^{\bfp}_1(n,k)]\geq \E [\nu_1(n,k)].$
\end{theorem}

\begin{proof} 
By (\ref{eq:gen_t_k}), which was proven originally in~\cite{FGT92}, we have that 
\begin{align}
\nonumber \E [\nu_1^\bfp(n,k) ] & = \sum_{q=0}^{k-1} \int_0^\infty [u^q] \prod_{i=1}^n \left( 1+u(e^{p_i v} -1 ) \right)e^{-v} dv\\ \nonumber
& = \sum_{q=0}^{k-1}\int_0^{\infty} e^{-nv}\left(\sum_{\substack{I \subseteq [n]\\ \abs{I} = q}} \prod_{i\in I} (e^{p_iv}-1)\right) dv \\ \label{general_p_t_1}
& = \int_0^{\infty} e^{-nv}\cdot \sum_{q=0}^{k-1}\left(\sum_{\substack{I \subseteq [n]\\ \abs{I} = q}} \prod_{i\in I} (e^{p_iv}-1)\right) dv.
\end{align}
Define $f(p_1,\ldots,p_n)\triangleq\sum_{q=0}^{k-1}\left(\sum_{\substack{I \subseteq [n]\\ \abs{I} = q}} \prod_{i\in I} (e^{p_iv}-1)\right)$. We show next that $f$ is minimized if and only if $p_i=\frac{1}{n}$ for all $1\le i\le n$. Furthermore, since $f$ is minimized if and only if $ \E [\nu_1^\bfp(n,k) ]$ is minimized, this concludes the proof.

Define $g(\bfp)=-1+\sum_{i=1}^np_i$. Using Lagrange multipliers, the Lagrangian function is
    \[
    \cL(\bfp,\lambda) = f(\bfp) + \lambda g(\bfp) = \sum_{q=0}^{k-1}\left(\sum_{\substack{I \subseteq [n]\\ \abs{I} = q}} \prod_{i\in I} (e^{p_iv}-1)\right) - \lambda + \lambda\sum_{i=1}^n p_i.
    \]
    We are looking for values of $\bfp$, that satisfy  
    \begin{align}
    \label{eq: partial d by lambda}
        \frac{\partial \cL(\bfp,\lambda)}{\partial \lambda} = -1 + \sum_{i=1}^n p_i = 0,
    \end{align}
    and for all $1\le i\le n$,
    \begin{align*}
        \frac{\partial \cL(\bfp,\lambda)}{\partial p_i} 
        & =
        \lambda + \sum_{q=1}^{k-1}\left(\sum_{\substack{I \subseteq [n]\backslash\{i\}\\ \abs{I} = q-1}} ve^{p_iv}\prod_{j\in I} (e^{p_jv}-1)\right) = 0,
    \end{align*}
    which is equivalent to
    \begin{align}
    \label{eq: partial d by pi}
        \lambda =  - ve^{p_iv} \sum_{\substack{I \subseteq [n]\backslash\{i\}\\ \abs{I} < k-1}} \prod_{j\in I} (e^{p_jv}-1).
    \end{align}
    Hence, for any $1\le i< i'\le n$ we have that 
    \[
    e^{p_iv} \sum_{\substack{I \subseteq [n]\backslash\{i\}\\ \abs{I} < k-1}} \prod_{j\in I} (e^{p_jv}-1) = 
    e^{p_{i'}v} \sum_{\substack{I \subseteq [n]\backslash\{i'\}\\ \abs{I} < k-1}} \prod_{j\in I} (e^{p_jv}-1).
    \]
    By reorganizing the latter equation, we have that for any $1\le i< i'\le n$,
    \[
    \left(e^{p_iv} - e^{p_{i'}v}\right) \sum_{\substack{I \subseteq [n]\backslash\{i,i'\}\\ \abs{I} < k-1}} \prod_{j\in I} (e^{p_jv}-1) =  \left(e^{p_iv} - e^{p_{i'}v}\right) \sum_{\substack{I \subseteq [n]\backslash\{i,i'\}\\ \abs{I} < k-2}} \prod_{j\in I} (e^{p_jv}-1),
    \]
    which is equivalent to 
    \[
    \left(e^{p_iv} - e^{p_{i'}v}\right) \sum_{\substack{I \subseteq [n]\backslash\{i,i'\}\\ \abs{I} = k-2}} \prod_{j\in I} (e^{p_jv}-1) = 0.
    \]
    Hence, we have that $p_i=p_{i'}$ or that $\abs{\{j: j\ne i,i', p_j>0\}} <k-2$. To complete the proof, let us show that the minimum is not attained for any $\bfp$ such that, $\abs{\mathsf{supp}(\bfp)} <n$.
    We prove the latter using an induction on $n$. For clarity, we will use the notation $f_n$ to indicate the relevant value of $n$ and $\bfp_{\bfu}^n \triangleq (\frac{1}{n}, \ldots, \frac{1}{n})$. The base case in which $n=2$ can be verified manually. This implies that $\bfp = \bfp_{\bfu}^2 =  \left( \frac{1}{2}, \frac{1}{2} \right)$ is the only minimum point for $f_2$.
    Assume the claim holds up to $n$, and let us prove its correctness for $n+1$. Let $\bfp=(p_1, \ldots, p_{n+1})$ be a minimum point for $f_{n+1}$, and assume by contradiction that $\abs{\mathsf{supp}(\bfp)} <n+1$ and further assume w.l.o.g. that $p_{n+1} = 0$.  Define $\bfp' = (p_1, \ldots, p_n)$ and note that $f_{n+1} (\bfp) = f_{n}(\bfp')$.  By the induction assumption, we know that a minimum point of $f_n$ has a support of size $n$ and hence, by the analysis of the Lagrangian function, we have that $f_n$ has a unique minimum point at $p_{\bfu}^n$. Therefore, we have that 
    $$
    f_{n+1} (\bfp) = f_{n} (\bfp') \ge f_n(\bfp_{\bfu}^{n} ),$$
    and equality is obtained if and only if $\bfp'=\bfp_{\bfu}^{n}$.
    Moreover, \autoref{cl:dec-function} implies that $f_n(\bfp_{\bfu}^{n})> f_{n+1}(\bfp_{\bfu}^{n+1})$, and thus, 
    $$
    f_{n+1} (\bfp)\ge f_n(\bfp_{\bfu}^{n}) > f_{n+1}(\bfp_{\bfu}^{n+1}),
    $$
   which is a contradiction. Thus, we get that $\abs{\mathsf{supp}(\bfp)} =n+1$, which implies that the only minimum point of $f_{n+1}$ is $\bfp_u^{n+1}.$

\end{proof}

\autoref{lem:uni-opt}, together with the previous claims imply a lower bound on $\E [\nu^{\bfp}_1(n,k)]$, which is given next.
\begin{corollary}
\label{th:MDS}
For any channel distribution $\bfp$ and any $(n,k)$ code $\cC$, it holds that, 
\begin{align} \label{eq:exp_uniform_optimality}
\E [\nu^{\bfp}_1(\cC)] \ge \E [\nu^{\bfp}_1(n,k)]\ge \E [\nu_1(n,k)]  = \sum_{i=0}^{k-1} \frac{n}{n-i},
\end{align} 
and if $\lim_{n \to \infty} n-k = \infty$ then $\sum_{i=0}^{k-1} \frac{n}{n-i}  \approx n \log(\frac{n}{n-k}) $.
Moreover (\ref{eq:exp_uniform_optimality}) holds with equality if and only if $\bfp = \bfp_u$.
\end{corollary}

Finally, we give the asymptotic value for the minimum expected sample size for the noiseless channel, $\E [\nu_1(n,k)]$. 

\begin{theorem} \label{th:optimal_ex}
Let $R$ be a constant, $0<R<1$. Then,  we have that $$\lim_{n \to \infty} \frac{\E[\nu_1(n,k =\lfloor nR\rfloor)]}{k} = \frac{1}{R} \log \left(\frac{1}{1-R} \right).$$
Furthermore, consider a sequence of MDS codes $\{\cC_i\}_{i=1}^\infty$ with parameters $[n_i,k_i]$ such that $\lim_{i\to\infty}k_i/n_i = 0$. Then, $$\lim_{i \to \infty} \frac{\E[\nu_1(n_i,k_i)]}{k_i} = 1.$$

\end{theorem}

\begin{proof}
If $0 < R < 1$ is fixed, then $n$ goes to infinity together with $k$ and thus we have that,   

\begin{align*}
& \lim_{n \to \infty} \frac{\E[\nu_1(n,k = \lfloor nR \rfloor) ] }{k} 
 = \lim_{n \to \infty} \frac{n(H_{n} -H_{n-k})}{k}  =   \frac{1}{R} \log \left(\dfrac{1}{1-R} \right),  
\end {align*} 
where the first equality holds from~\cite{Feller} and the second equality is a known result.

Furthermore, in the case in which we have a sequence of MDS codes $\{\cC_i\}_{i=1}^\infty$, such that $\lim_{i \to \infty} \frac{k_i}{n_i} =0,$ the equality below holds. $$\lim_{i \to \infty} \frac{\E [\nu_1(n_i,k_i)]}{k_i} = \lim_{i \to \infty} \frac{n_i(H_{n_i} -H_{n_i-k_i})}{k_i} = \lim_{i \to \infty} \frac{\sum_{j=0}^{k_i-1} \frac{n_i}{n_i-j}}{k_i},$$
where, 
\begin{align*}
   \lim_{i \to \infty} \frac{\sum_{j=0}^{k_i-1} \frac{n_i}{n_i-j}}{k_i} \le \lim_{i \to \infty} \frac{k_i \left( \frac{n_i}{n_i-(k_i-1)}\right)}{k_i} = \lim_{i \to \infty} \frac{k_i \left( \frac{1}{1-\frac{k_i-1}{n_i}}\right)}{k_i} = 1,
\end{align*}
and, 
\begin{align*}
    \lim_{i \to \infty} \frac{\sum_{j=0}^{k-1} \frac{n_i}{n_i-j}}{k_i} \ge \lim_{i \to \infty} \frac{k_i \left(\frac{n_i}{n_i-0}\right)}{k_i} =1.
\end{align*}
Thus, we can conclude that, 
$\lim_{i \to \infty} \frac{\E [\nu_1(n_i,k_i)]}{k_i} =1$.
\end{proof}

\section{The MDS Coverage Depth Problem - The Noisy Channel} 
\label{sec:sequencing coverage:erroneous}
The main goal of this section is to address Problem~\ref{pr:expectation_prob} for the noisy channel under the uniform distribution. Under this setup, we assume the data is encoded with an $[n,k]$ MDS code and that each strand $\bfx_i$ can be retrieved given some $t>1$ reads, which are noisy copies of it, and cannot be retrieved given less than $t$ reads. 
Similarly to the previous section, it is enough to successfully decode $k$ (or more) sequences $\bfx_i$ in order to retrieve the stored information and so under this setup the minimum sample size for our problem is equivalent to the quantity $\nu_{t}(n,k)$ where $t > 1$.

It should be noted that as listed in the related work section, the first part of \autoref{pr:expectation_prob}, i.e., the value of $\E [\nu_{t}(n,k)]$ is known~\cite{FGT92} and is given in (\ref{eq:gen_t_k}). However, it is not a closed-form expression, and in this section, we give several closed-form expressions that bound this value and thus extend the known result. Furthermore, the most related result regarding the probability distribution $P[\nu_t(n,k)>m]$ was given in~\cite{ErdosReyni}. The authors showed that for $n=k$ any $x \in \R $, the probability satisfies $P[\nu_t(n,n)>n\log n +(t-1)\log \log n + n x] \le e^{-\frac{e^{-x}}{(t-1)!}}$. 

In this section, we extend the latter result, by providing several bounds for the case when $k<n$, which is assumed for the rest of this section. Our main results for this case are stated in \autoref{th:concentrate} and in \autoref{cl:lower_bound_prob}. To discuss these results, we first define the following value. Given $n$, $k$, and $t$ as stated above, 
we define
\begin{align} \label{eq:r(n,k,t)}
r(n,k,t) \triangleq n \log \left( \frac{n}{n-k} \right) + n t \log \log n + 2 n \log (t+1). 
\end{align}
In \autoref{th:concentrate}, it is shown that when $n$ is large enough, the probability that more than $r(n,k,t) $ reads are required to retrieve the information i.e., $ P[\nu_{t} (n,k) > r(n,k,t) ]$  approaches zero.

Furthermore, for the case in which $k = Rn$, where $0<R<1$ is a fixed constant, the value of $r(n,k,t)$ can be reduced by replacing the expression $\log \log (n)$ with any function of $n$ that approaches to infinity with $n$. To conclude this discussion, we also show in \autoref{cl:lower_bound_prob} that for any $c$, the probability that less than $n \log(\frac{n}{n-k}) -nc$ reads are enough to retrieve the information is bounded from above by $e^{-c} (1 + \frac{1}{n-k})$.  
We start by showing that for any $\varepsilon>0$, $P \left [ {\nu_{t}(n,k)}  \leq  {r(n,k,t) } \right ] \ge 1- \varepsilon$ for $n$ large enough. 

\begin{theorem}\label{th:concentrate}
 For any $\varepsilon$ and $n$, such that  $\varepsilon>0$, $n>e^{\frac{6t\cdot 2^{t-1}}{\varepsilon}} \ge 16$,  we have that,
\begin{align*}
P \left [ {\nu_{t}(n,k)}  \leq  {r(n,k,t) } \right ] \ge 1- \varepsilon
\end{align*}
\end{theorem} 

\begin{proof}
To prove the statement in the theorem it is suffice to show that $P \left [ {\nu_{t}(n,k)}  > {r(n,k,t) } \right ] < \varepsilon.$ 
Denote $r \triangleq r(n,k,t)$ and recall that within the context of the urn problem (see \autoref{sec:sequencing coverage:noiseless}), the random variable $\nu_{t}(n,k)$ denotes the number of balls (or rounds) until we have a set of $k$ urns where each urn has at least $t$ balls. Hence, we show that if the number of balls thrown is at least $r$, then the probability of having $n-k+1$ or more urns which are \emph{not} filled with $t$ balls is approaching zero.
The approach leveraged in the proof is inspired by a technique first employed by Erd\H{o}s and R\'{e}nyi in~\cite{ErdosReyni}. 
Let us define the following event.
\begin{description}
\item[$E^{(r)}_{t}$:]  After $r$ rounds, there exists a set $S_{t}$, of $n-k+1$ urns, each containing less than $t$ balls. 
\end{description}
Next, we show that the probability of $E^{(r)}_{t}$ approaches zero when $n$ is large. To this end, we define $z_i(n,r)$ for $1\le i \le n$, as a random variable that governs the number of balls in the $i$-th urn, after $r$ draws.
For $n$ large enough, the probability that urn $i$ has at most $t-1$ balls after $r$ draws is denoted by $P[z_i(n,r) \le t-1]$ 
and is given by, 
\begin{align*}
P[z_i(n,r) \le t -1 ] 
&= \sum_{j=0}^{t-1}\binom{r}{j} \left( \frac{1}{n} \right)^{j} \left( 1- \frac{1}{n} \right)^{r-j}  
\\ & {\le t \cdot \binom{r}{t-1} \left( \frac{1}{n} \right)^{t-1} \left( 1-\frac{1}{n} \right)^{r -(t-1)} }
\\& \le t \cdot \left( \frac{r\cdot  e}{t-1} \right)^{t-1} \left( \frac{1}{n} \right)^{t-1} \left( 1-\frac{1}{n} \right)^{r -(t-1)},
\end{align*}  
where the first inequality is proven in \autoref{claim:binomial} in Appendix~\ref{Appendix_ExpectationBound_f}, and the last inequality follows from the fact that $\binom{r}{t-1} \le (\frac{re}{t-1})^{t-1}$. 
Note that ${(\frac{e}{t-1}})^{t-1} <3$, for $t>1$. Thus, 
\begin{align*}
P[z_i(n,r) \le t -1 ]
& \le 3  t \cdot \left( \frac{r}{n} \right)^{t-1}  \left( 1-\frac{1}{n} \right)^{n \left( \frac{r}{n} -\frac{t-1}{n} \right)}.
\end{align*}  
We have that, 
\begin{align*}
P \left[z_i(n,r)\le t -1 \right] & \le 3 t \cdot \left( \log \left( \frac{n}{n-k} \right) + t \log \log (n) +  2 \log(t+1) \right)^{t-1} \left(e^{\left( \frac{-r}{n} +\frac{t-1}{n} \right)} \right)
\\ & \le 3 t \cdot (2 \log n)^{t-1} \left(\frac{n-k}{n} \right) \left( \frac{1}{\log^t n}\right) \left( \frac{1}{(t+1)^2}\right) e^{\frac{t-1}{n}}
\\ &= 3t \cdot  \frac{e^{\frac{t-1}{n}}}{(t+1)^2}\cdot   \frac{(2 \log n)^{t-1}}{\log^t(n)} \cdot \frac{n-k}{n}  
\\ &= 3t \cdot  \frac{e^{\frac{t-1}{n}}}{(t+1)^2}   \cdot \frac{2^{t-1}}{\log(n)} \cdot \frac{n-k}{n}, 
\end{align*}
where the second inequality holds since for $n$ large enough  $\log(\frac{n}{n-k}) + t \log \log (n) + 2 \log (t+1) \le (2 \log n)$.
It should be noted that for $n>t$, which is the case of our interests,  we have that  $3 t \cdot  \frac{e^{\frac{t-1}{n}}}{(t+1)^2}  \le 6t$, and hence, 

$$ P \left[z_i(n,r)\le t -1 \right] \le 6t \cdot \frac{2^{t-1}}{\log(n)} \cdot \frac{n-k}{n}.$$
Now let us define a random variable $Y$ as the number of urns with less than $t$ balls. From the linearity of expectation, regardless if the urns are independent or not, the expected number of urns that have less than $t$ balls is, 
\begin{align*}
\E [Y] & = \Sigma_{i=1}^n \E[z_i(n,r)] 
\\ & =  n  \cdot P\left[ z_i(n,r)\le t-1 \right] \le   (n-k) \cdot 6t \cdot \frac{2^{t-1}}{\log(n)},
\end{align*}
where the last inequality holds for $n$ large enough. 

Note that 
$$P\left[E_{t}^{(r)}\right] = P[Y \ge n -k +1],$$
and hence by Markov's inequality, we can conclude that, 
$$P[Y \ge n -k +1 ] \le \frac{\E[Y]}{n-k+1} < 6t \cdot \frac{2^{t-1}}{\log(n)}. $$
Thus, we get that $P[E_{t}^{(r)}] \to 0$ for $n$ large enough which implies the statement in the theorem. 
\end{proof}
For fixed-rate codes, i.e., for the case where $k = Rn$, when $0<R<1$, and $R$ is a fixed constant (when $n$ grows), we present a stronger result in the next theorem. The proof of this theorem can be found in Appendix \ref{Appendix_ExpectationBound_f}.

\begin{theorem}\label{th:concentrate_fixed}
Let $f: \N \to \R$ be a function such that $\lim_{n\to \infty} f(n) =\infty$, and let  
\begin{align} \label{eq:r_f}
 r_f(n,k=Rn,t) \triangleq n \log \left(\frac{1}{1-R}\right) +n t f(n) + 2n (t+1).
\end{align}
Then, for $n$ large enough, it holds that
\begin{align*}
P \left [ {\nu_{t}(n,k)}  > {r_f(n,k,t)} \right ] \le 6 t^t \frac{(2 \cdot f(n))^{t-1}}{e^{t \cdot f(n)}} \cdot \left(1-R \right) . 
\end{align*}
\end{theorem}

\autoref{th:concentrate_fixed} draws a connection between the sample size and the probability of successful retrieval when using fixed-rate codes. In particular, using the results of \autoref{th:concentrate_fixed} one can pick any function $f(n)$ that approaches infinity as slowly (or fast) as possible to get an upper bound on this probability which gets bigger (or smaller). 

Next, for any $c\in\R$,  we denote, 
\begin{align} \label{eq:r_L}
r_L (n,k, c) \triangleq  n \log \left(\frac{n}{n-k}\right) -nc.
\end{align}
In the next lemma, an upper bound on the probability $P[\nu_t(n,k) \le  r_L(n,k,c)]$ is given.

\begin{lemma} \label{cl:lower_bound_prob} For any $c>0$, and any $t\ge 1$ it holds that, 
$$P\left[ \nu_t(n,k) \le  n \log (\frac{n}{n-k}) -nc \right] \le e^{-c} \left(1+\frac{1}{n-k}\right).$$
\end{lemma}
\begin{proof}
 We first highlight that $\nu_t(n,k) \ge \nu_1 (n,k)$, and thus it is enough to show that $$P\left[ \nu_1(n,k) \le  n \log (\frac{n}{n-k}) -nc \right] \le e^{-c} \left(1+\frac{1}{n-k} \right).$$
We have that, 
    \begin{align*}
        e^{\log (\frac{n}{n-k}) -c} \cdot \E \left[ e^{-\frac{\nu_1 (n,k)}{n} } \right] &= e^{\log (\frac{n}{n-k}) -c} \cdot \sum_{j=1}^\infty e^{-\frac{j}{n}} P\left[ {\nu_1 (n,k)} =j \right] 
        \\ & = \sum_{j=1}^\infty e^{\log (\frac{n}{n-k}) -c-\frac{j}{n}}  P\left[ {\nu_1 (n,k)} =j \right] 
        \\ & = \hspace{-2ex} \sum_{j=1}^{\lfloor n\log (\frac{n}{n-k}) -nc \rfloor} \hspace{-3ex} e^{\log (\frac{n}{n-k}) -c-\frac{j}{n}}  P\left[ {\nu_1 (n,k)} =j \right] + \hspace{-5ex} \sum_{j=1+\lfloor n \log (\frac{n}{n-k}) -nc \rfloor}^{\infty} \hspace{-5ex} e^{\log (\frac{n}{n-k}) -c-\frac{j}{n}}  P\left[ {\nu_1 (n,k)} =j \right]
        \\ & \ge \sum_{j=1}^{\lfloor n\log (\frac{n}{n-k}) -nc \rfloor} \hspace{-3ex} e^{\log (\frac{n}{n-k}) -c-\frac{j}{n}}  P\left[ {\nu_1 (n,k)} =j \right]
        \\ & \ge \sum_{j=1}^{\lfloor n\log (\frac{n}{n-k}) -nc \rfloor} \hspace{-3ex} 1 \cdot  P\left[ {\nu_1 (n,k)} =j \right]
        \\ & \ge P\left[ {\nu_1 (n,k)} \le n \log \left[ \frac{n}{n-k} \right] - nc \right].
    \end{align*}

    {From~\cite{PG83}, the  generating function of the geometric random variable $\nu_1(n,k)$
    is given by} 
    \begin{align*}
    G_{\nu_1(n,k)} (x) = \E [x^{\nu_1(n,k)}] = \sum_{j=0}^\infty P[\nu_1(n,k)=j]x^j 
 =\prod_{i=1}^{k} \frac{(n-{(i-1)})x}{n-{(i-1)x}}  =\prod_{i=1}^{k} \frac{(1-\frac{i-1}{n})x}{1-\frac{i-1}{n}x}.
     \end{align*}

Thus, given $x=e^{-1/n}$, we get that, 
    \begin{align*}
     \E \left[ e^{\frac{-\nu_1 (n,k)}{n} } \right] &=  \prod_{i=1}^{k}  \frac{(1-\frac{i-1}{n}) e^{-\frac{1}{n}}}{1-(\frac{i-1}{n})e^{-\frac{1}{n}}}
    \\ &=    \prod_{i=1}^{k} \frac{1-\frac{i-1}{n}}{e^{\frac{1}{n}}- \frac{i-1}{n} } \le   \prod_{i=1}^{k} \frac{1-\frac{i-1}{n}}{1+\frac{1}{n}  - \frac{i-1}{n}} 
    \\ & =   \prod_{i=1}^{k} \frac{1-\frac{i-1}{n}}{1 - \frac{i-2}{n} } =   \frac{1-\frac{k-1}{n} }{1+\frac{1}{n} } = \frac{n-k+1}{n+1} \le \frac{n-k+1}{n},
    \end{align*}
    where in the first inequality we used the fact the $e^{\frac{1}{n}} \ge 1 + \frac{1}{n}.$
Hence, it holds that, for positive $c$,
\begin{align*}
    e^{\log(\frac{n}{n-k}) -c} \cdot  \E \left[ e^{-\nu_1(n,k)/n} \right] \le e^{-c} \cdot \frac{n}{n-k} \cdot  \frac {n-k+1}{n} = e^{-c} \left(1+\frac{1}{n-k}\right).
\end{align*}
Finally, we conclude that, 
\begin{align*}
P\left[ \nu_1(n,k) \le   n \log (\frac{n}{n-k}) -nc \right] \le e^{\log(\frac{n}{n-k}) -c} \cdot \E \left[ e^{-\nu_1(n,k)/n} \right] \le  e^{-c}\left(1+\frac{1}{n-k}\right).
\end{align*}
\end{proof}


Combining \autoref{cl:lower_bound_prob} and \autoref{th:concentrate}, and assuming $t$ is a constant with respect to $n$, in the next theorem we show upper and lower bounds on   $\E \left[\frac{\nu_t (n,k)}{n} \right]$. 
\begin{theorem} \label{th:lower_upper_exp}
For any $\varepsilon >0$, there exists $n_{\varepsilon}$, such that for any $n>n_{\varepsilon}$ we have that, 

\vspace{-1ex}
\begin{small}
$$  \log\left(\frac{1}{1-R} \right) +f_c(n,R)  \le \E \left[ \frac{\nu_t (n,k)}{n} \right] \le  \left(  \log \left(\frac{1}{1-R}\right) +  t \log \log n + 2  \log (t+1) \right)\cdot (1+ 2\varepsilon ) , $$   
\end{small}
where $f_c(n,R)=\frac{1}{2n}( 1-\frac{1}{1-R}) -  \sum_{h=1}^{\infty} \frac{B_{2h}}{2hn^{2h}} \left(1- \frac{1}{(1-R)^{2h}} \right) = \cO(\frac{1}{n^2}),$ and $B_h$ denotes the $h$-th Bernoulli number.
\end{theorem}

\begin{proof}
First, we highlight that for any integer $t>0$, it holds that $\nu_t(n,k) \ge \nu_1(n,k)$. Next, we recall the known results proven in~\cite{FGT92}, where they showed $\E[\nu_1(n,k)] = n (H_n - H_{n-k})$. Hence, we can conclude that the following holds for $n$ large enough, 
\begin{align*}
\E[\nu_t(n,k)] &\ge \E[\nu_1(n,k)]
                \\ & = n(H_n -H_{n-k}) 
                \\ & = n \left(\log(n) + \gamma + \frac{1}{2n} - \sum_{h=1}^{\infty} \frac{B_{2h}}{2hn^{2h}} - \log(n-k) - \gamma - \frac{1}{2(n-k)} + \sum_{h=1}^{\infty} \frac{B_{2h}}{2h(n-k)^{2h}} \right) 
                \\ & = n \log\left( \frac{n}{n-k}\right) +  \frac{1}{2}\left( 1-\frac{1}{1-R} \right) - n \sum_{h=1}^{\infty} \frac{B_{2h}}{2hn^{2h}} \left(1- \frac{1}{(1-R)^{2h}} \right)
                \\ & =  n \log\left(\frac{1}{1-R}\right) +nf_c(n, R), 
\end{align*}
where  $\gamma \sim 0.5772156649 $ is the Euler-Mascheroni constant, where the last equality was proven {in~\cite{FGT92}}.
Next, let $r_n\triangleq r(n,k,t)$ (recall that by (\ref{eq:r(n,k,t)}), $r(n,k,t) = n\log(\frac{1}{1-R}) +nt \log \log (n) + 2 n \log(t+1)$). In \autoref{th:concentrate} we showed that $P[\nu_t(n,k) > r_n] < \varepsilon$. Using the same methods, in Appendix~\ref{appendix:proof_of_expectation} we proved \autoref{th:concentrate_for_exp} which states that for any integer {$i \ge 1$} and for $n$ large enough,  $P[\nu_t(n,k) > r_n \cdot i]  < \varepsilon \cdot \frac{i^{t-1}}{\log^{t(i-1)}(n)} ,$ and thus we can conclude that, 
\begin{align*}
E[\nu_t(n,k)] & = \sum_{r\in\N} P(\nu_t(n,k) \ge r )
\\ & = \sum_{r<r_n} P(\nu_t(n,k) \ge r ) + \sum_{r\ge r_n} P(\nu_t(n,k) \ge r )
\\ &  \le 1 \cdot r_n +\sum_{r\ge r_n} P(\nu_t(n,k) \ge r )
\\ &  =   r_n +\sum_{i=1}^{\infty} \sum_{r=i\cdot r_n}^{(i+1)\cdot r_n}  P(\nu_t(n,k) \ge r ) 
\\ & \le  r_n +\sum_{i=1}^{\infty} \sum_{r=i\cdot r_n}^{(i+1)\cdot r_n}  P(\nu_t(n,k) \ge i \cdot r_n ) 
\\ &  = r_n +\sum_{i=1}^{\infty} r_n \cdot  P(\nu_t(n,k) \ge i \cdot r_n ) 
\\ &  <  r_n +\sum_{i=1}^{\infty} r_n \cdot \varepsilon \cdot \frac{i^{t-1}}{\log^{t(i-1)}(n)} 
\\ &  =   r_n +\varepsilon \cdot r_n \sum_{i=1}^{\infty}    \frac{i^{t-1}}{\log^{t(i-1)}(n)}  \\ & \stackrel{(a)}{<} r_n + 2 \varepsilon\cdot  r_n 
\\ & =  r_n\cdot (1+ 2\varepsilon ),
\end{align*}
where (a) follows since $\sum_{i=1}^{\infty}    \frac{i^{t-1}}{\log^{t(i-1)}(n)} <2$ for $n$ large enough and any integer $t>0$.
Lastly, we simplify the expression, 

\begin{align*}
    \frac{1}{n} r_n(1+2\varepsilon) & = \left(  \log \left(\frac{1}{1-R}\right) +  t \log \log n + 2  \log (t+1) \right)\cdot (1+ 2\varepsilon )  \\ & = \log\left(\frac{1}{1-R}\right) + \cO(t\log\log n),
\end{align*} 
which completes the proof. 
\end{proof}

For practical purposes of DNA storage systems, it is sometimes required to plan ahead and sample the number of reads that guarantees successful decoding with high probability. Hence, we turn to the following strongly related problem and give a closed-form expression to the corresponding value. Turning back to the urn problem terminology, we define $X^{(r)}$ as the number of urns that are not filled with at least $t$ balls after $r$ rounds. The goal is to find a lower bound on the number of rounds $r$, that guarantees that the expected number of urns that are \emph{not filled} with $t$ balls is at most $n-k$. That is, to find $r_E$, such that for any $r\ge r_E$, we have that $\E [X^{(r)}] \le n-k$. In order to derive this result, we first consider the probability that any fixed urn is \emph{not filled} with $t$ or more balls by the $r$-th round. This probability is given by, 
$$p = \sum_{j=0}^{t-1} \binom{r}{j} n^{-j} \left(1-\frac{1}{n}\right)^{r-j} \leq e^{-rD(\frac{t-1}{r} || \frac{1}{n})},$$
where the last inequality follows from Chernoff bound~\cite{C52} for $r\ge n(t-1)$, and $D(a|| p)$ is the Kullback–Leibler divergence~\cite{C75} which is given by 
$$
D(a||p) \triangleq a\log_2\frac{a}{p} + (1-a)\log_2\frac{1-a}{1-p}.
$$

Under our setup, each of the $n$ urns can be interpreted as a Bernoulli random variable with probability $p$, which is denoted by $X^{(r)}_i$ for $1 \le i \le n$. Note that $X^{(r)} = \sum_{i=1}^n X^{(r)}_i$ is the number of urns that are not filled with at least $t$ balls after $r$ rounds, which implies that the number of urns that have at least $t$ balls is $n - X^{(r)}$. Our approach will be to determine a value for $r$, which guarantees (in expectation) that $X^{(r)}$ is at most $n-k$. From the linearity of expectation,
\begin{align}\label{eq: upper bound on E[X(r)]}
    \E [X^{(r)}] = np \le  n  e^{-(t-1) \log_2(\frac{n(t-1)}{r})- (r-(t-1)) \log_2 \left(\frac{(r-(t-1))n}{r(n-1)}\right)}.
\end{align}

The next claim will be used in the derivation to follow and its proof can be found in Appendix \ref{Appendix_LowerBound_r}.
\begin{claim}\label{claim::condition for E[X]<=n-k}
For $r\ge n(t-1)$, we have that $\E [X^{(r)}] \le n-k$, if, 
\begin{align}\label{eq: sufficient cond on r and R}
-\frac{r}{n(t-1)} e^{-\frac{r}{n(t-1)}} \ge -\frac{1}{e} \left(1-\frac{k}{n}\right)^{\frac{\log 2}{t-1}}.
\end{align}
\end{claim}
 Using known results on the Lambert W function{~\cite[Section IV]{CG96},~\cite[Theorem 1]{C13}}, the values of $r$ for which (\ref{eq: sufficient cond on r and R}) holds can be concluded. This is summarized in the next theorem, and the complete proof can be found in Appendix \ref{Appendix_LowerBound_r}. For $0<R<1$, we denote, 
 \begin{align} \label{eq:r_E}
 r_E(n, k=Rn, t) \triangleq n(t-1)-n\log2 \log(1-R) + n(t-1)\sqrt{-\frac{2\log2}{t-1}\log(1-R)}.
 \end{align}

\begin{theorem}\label{the: lower bound r}
Let $R=\frac{k}{n}$. For any $r \ge r_E(n, k, t),$
we have that  $\E [X^{(r)}] \le n-k$. 
\end{theorem}

At this point, we would like to shed some light on the relation between~\autoref{th:concentrate}, \autoref{th:concentrate_fixed} and~\autoref{the: lower bound r}. 
In our setup, which uses the urn problem terminology, it is assumed that $r$ balls are thrown into $n$ unique urns, and we are interested in the event that at least $k$ of these urns contain at least $t$ balls each. This scenario can be parameterized in two different ways; (a) the number of balls that need to be thrown, and (b) the number of urns that contain $t-1$ or less balls. The random variable $\nu_{t}(n,k)$ governs the value in (a), assuming that the value in (b) is fixed. Analogously, the random variable $X^{(r)}$ governs the value in (b), assuming that the value in (a) is fixed. 

\begin{figure}
\centering
\begin{subfigure}{.95\textwidth}
  \centering
  \includegraphics[width=\linewidth]{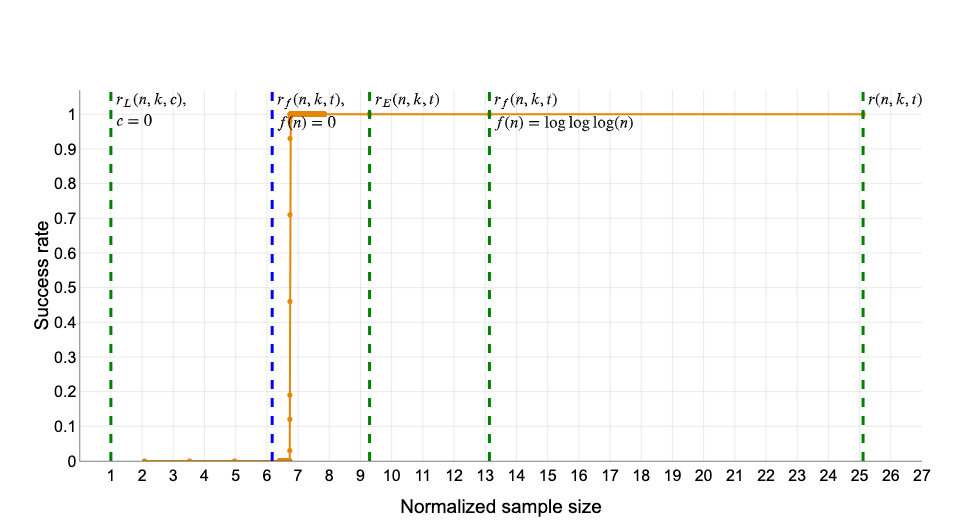}
  \caption{{$R=0.5$}}
  \label{fig:sub1}
\end{subfigure}%

\begin{subfigure}{.95\textwidth}
  \centering
  \includegraphics[width=\linewidth]{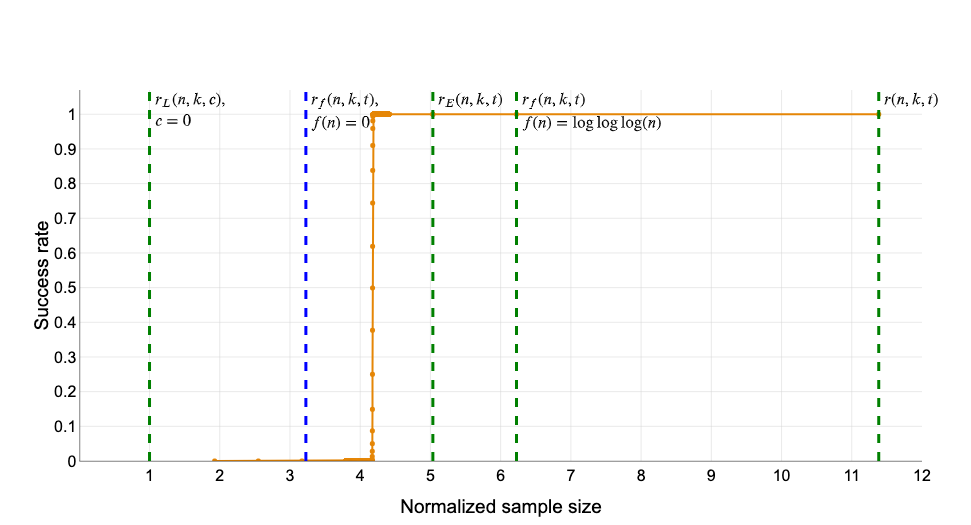}
  \caption{$R=0.8$}
  \label{fig:sub2}
\end{subfigure}
\caption{Simulation results of the success rate (fraction of successful experiments) as a function of the number of draws. The X-axis shows the number of draws (normalized by $n\log(\frac{1}{1-R})$), while the Y-axis shows the fraction of simulations in which there were at least $k$ urns with $t$ balls each. The parameters used in the simulations were $n=100,000$, $t=5$, and for each number of draws, we had $1,000$ simulations. It can be seen that for both~\autoref{th:concentrate} and \autoref{the: lower bound r} the success rate of $1$, as expected.}
\label{fig:simulation_expected}
\end{figure}

In the case where the probability distribution is tightly concentrated (i.e., where $\nu_{t}(n,k)$ is tightly concentrated around its mean and similarly for $X^{(r)}$), one would expect these two quantities to coincide. 
\autoref{fig:simulation_expected}, shows results from computer simulations we made to demonstrate the results of \autoref{th:concentrate} and \autoref{the: lower bound r}. In the presented simulation we used $n=100,000$ urns, $R \in \{ 0.5, 0.8\}$, $k=Rn$, and $t=5$. In each simulation $r$ balls are drawn, each inserted into one of the urns randomly, and the simulation is considered as success if it ends with at least $k$ urns, each with at least $t$ balls. For any value of $r$, the presented result is the fraction of successful simulations out of $1,000$ simulations we have made per $r$. The Y-axis shows the fraction of successful experiments, and the X-axis shows the number of draws $r$ normalized by $n \log (\frac{1}{1-R})$. It can be seen that the success rate of both values presented in \autoref{th:concentrate} ($r(n,k,t)$) and \autoref{the: lower bound r} ($r_E(n,k,t)$) are $1$. 

Practically speaking, as mentioned above, the noisy channel fits the real scenario of DNA storage systems. Hence, it should be mentioned that a similar problem was studied experimentally by Erlich and Zielinski~\cite{EZ17}, however, with a slightly different setup. They presented the DNA fountain, a Luby transform-based scheme and assumed that the total number of reads is fixed and is given (from the DNA sequencer) and it is distributed with a negative binomial distribution. Thus, they were able to calculate the average number of copies per strand and empirically evaluate the required sample size as a function of the distribution's parameters. It should be noted that they only considered reads of the design length and thus the error rates were reduced. They also evaluated how dilution affects the distribution and the required sample size. 

Finally, another variation of the noisy channel $\cS$ is studied, which is relevant to the DNA fountain~\cite{EZ17} and similar schemes. Here, it is required to obtain a single noiseless copy from $k$ out of the $n$ synthesized strands. Assuming uniform distribution on the strands, in this channel, any sampled read is drawn noiseless with some fixed probability $0< \alpha < 1$. We use the notation of $\omega_\alpha(n,k)$ to denote the random variable describing the required sample size to ensure successful decoding in this case. We note that this setup is easier to analyze, and the following results can be derived using similar techniques as in the classical coupons collector's problem~\cite{Feller}; see Appendix~\ref{appendix_noisy_alpha}.

\begin{theorem} \label{th:noisy_alpha}
For any $k\le n$, ${\E[\omega_\alpha(n,k)] = \frac{n}{\alpha} \left( H_n - H_{n-k} \right).}$ 
\end{theorem}

\section{Random Access}\label{sec:random}
In this section we study the problem of optimizing the sample size for random access queries in DNA storage systems. Recall that, in this problem, a vector of $k$ information strands each of length~$\ell$, $\bfU = ( \bfu_1, \bfu_2, \ldots, \bfu_k )\in (\Sigma^\ell)^k$, is encoded into a vector of $n$ strands, each of length $\ell$, 
$\bfX = (\bfx_1, \bfx_2, \ldots, \bfx_n) \in (\Sigma^\ell)^n$ that are stored in the DNA storage channel as described in \autoref{sec:defs}. Later, the user wishes to retrieve a single information strand $\bfu_i$ for some $i \in [k]$. Unless stated otherwise, we assume the channel is uniform and noiseless. 
 
We start by studying the case when the number of information strands matches the number of coded strands (i.e. $n=k$), and prove that the optimal retrieval strategy involves no coding, resulting in an expected retrieval time of $k$. Next, we extend our insights to more involved cases, including systematic MDS codes, affirming that the expected retrieval time remains $k$ for any information strand. Finally, we present explicit code constructions achieving expected retrieval times below $k$ and evaluate their performance analytically and through simulations, while also providing lower bounds on the maximum expected retrieval time in different scenarios. 

\subsection{Preliminary Results}
Recall that given an $(n,k)$ code $\cC$, for ${i\in [k]}$, we denote by $\tau_i(\cC)$ the random variable that governs the number of samples to recover the $i$-th information strand. The next lemma fully solves \autoref{prob:random:single} when no coding is used. 

\begin{lemma} 
\label{lem:random:single:no code}
Let $n\ge 1$. For any $1 \le i \le n$, we have that 
\begin{enumerate}
\item $\E[\tau_i] = n$ and  $T_{\max}=T_{\textrm{avg}}=n.$  
\item For any $r\in\mathbb{N}$ we have that 
$P[\tau_i>r] =  \left( 1 - \frac{1}{n} \right)^r$, 
and
$P[\tau_i = r] =   \frac{1}{n} \cdot \left( 1 - \frac{1}{n} \right)^{r-1}$.
\end{enumerate}
\end{lemma}
\begin{proof}
For the first part, note that for any $i$, $\tau_i$ has geometric distribution with success probability $p=\frac{1}{n}$ and hence we have that $\E[\tau_i] =p^{-1} =n$  which implies that
$$T_{\max} = \max_{1\le i\le n} \mathbb{E}[\tau_i] =  p^{-1}=n,$$ 
and 
$$T_{\textrm{avg}} =  \frac{1}{n}\sum_ {i=1}^{n} \mathbb{E}[\tau_i] = \frac{np^{-1}}{n} =p^{-1}= n.$$

For the second part we have that $\tau_i > r$ for an integer $r$ only if $\bfu_i$ was not sampled in the first $r$ trials, and hence
\begin{align*}
P[\tau_i>r] =  (1-p)^r =\left( 1 - \frac{1}{n} \right)^r,
\end{align*}
and 
\begin{align*}
P[\tau_i = r] =  \frac{1}{n} \cdot \left( 1 - \frac{1}{n} \right)^{r-1}.
\end{align*}
\end{proof}

Before we continue with the analysis of more involved cases, we define the $n$ random variables $\widehat{\tau}_i(\cC), i\in [n]$, such that $\widehat{\tau}_i(\cC)$ governs the required sample size to retrieve the $i$-th encoded strand. Additionally, for every set $J\subseteq[n]$, let $\widehat{\tau}_J(\cC)\triangleq \max_{i\in J} \widehat{\tau_i}(\cC).$ These random variables are used as a technical tool in our analysis and the key idea is given in the next lemma. 
The proof follows the same ideas as the proof for the coupon collector's problem and is given here for completeness.

\begin{claim}
\label{claim:random:single:expected maximum retrieval time of encoded strands' subset}
For any $(n,k)$ code $\cC$ and any $J\subseteq[n]$ of size~$\rho$ we have that $ \mathbb{E}[\widehat{\tau}_{J}(\cC)] = nH_{\rho}.$
\end{claim}

\begin{proof}
Let $t_i$ for $1\le i\le \rho$ be the number of draws to collect the $i$-th strand in $J$ after the $(i-1)$-th strand from $J$ was collected. Note that $\widehat{\tau}_{J}(\cC) = \sum_{i=1}^{\rho}t_i$. Additionally, observe that $t_i$ is a geometric random variable with success probability $p_i=\frac{\rho-{i+1}}{n}$ and $\mathbb{E}[t_i]=\frac{1}{p_i}$. Hence, by the linearity of the expectation we have that 
\begin{align*}
\mathbb{E}\left[\widehat{\tau}_{J}(\cC)\right] &  = \mathbb{E}\left[{\sum_{i=1}^{\rho}t_i}\right] = \sum_{i=1}^{\rho}\mathbb{E}\left[t_i\right] = \sum_{i=1}^{\rho}\frac{n}{\rho-i+1} = n\sum_{i=1}^{\rho}\frac{1}{i} = nH_{\rho}.
\end{align*} 
\end{proof}

For the rest of this section, it is assumed that $\cC$ is an $(n,k)$ code and $\bfX$ is the encoded codeword of the information vector $\bfU$. 
The structure of $\cC$ defines for each information strand all the possible sets of encoded strands that are sufficient for its recovery. This concept is similar to recovery sets in \emph{locally repairable codes}~\cite{PD14} as well as the ones with \emph{availability}~\cite{FVY15,HY15,VY23}.

This can be defined formally as follows.

\begin{definition}
Let $\cC$ be an $(n,k)$ code. We say that $J\subseteq [n]$ is a \emph{retrieval set} of the $i$-th information strand (i.e., $\bfu_i$) if it is possible to decode the information strand $\bfu_i$ from the encoded strands whose indices belong to $J$. The set of all retrieval sets of $\bfu_i$ is denoted by $\widehat{\cD}(i)$, and $\cD(i)$ is the set of all minimal retrieval sets of $\bfu_i$ (with respect to the inclusion relation).
\end{definition}
We say that an $(n,k)$ code $\cC$ is a \emph{systematic} code if for any $i\in[k]$ 
it holds that $\bfu_i$ has a retrieval set of size one. In other words, $\cC$ is systematic if for any $i\in[k]$ we have that $$\min\{|J| : J\in \cD(i)\} = 1.$$ 

Next, we consider the case of non-systematic codes for $k=n$ (in particular $\bfU \neq \bfX$). Since $\bfX$ and $\bfU$
have the same length, given any set of strands $\{ \bfx_i : i \in J \}$, we can recover at most $|J|$ information strands from $\bfU$. Our goal is to extend \autoref{lem:random:single:no code} to the coded case when $k=n$ using this basic insight. 

\begin{claim} \label{cl:ra:n=k_nocode}
For any code $(n=k, k)$ $\cC$, we have that $T_{\max}^\cC\ge T_{\max} = n$ and $T_{\text{avg}}^\cC\ge T_{\text{avg}} = n$, where equality is obtained if and only if $\cC$ is systematic. 
In particular, if we let $\rho_i$ be the size of the smallest retrieval set for the information strand $\bfu_i$, then 
\begin{enumerate}
    \item $\E [ \tau_i(\cC)]=nH_{\rho_i}$,
    \item $T_{\max}^{\cC} = n H_\rho$,  where $\rho\triangleq \max_i \rho_i$,
    \item $T_{\text{avg}}^{\cC} = \sum_{i=1}^n H_{\rho_i}.$ 
\end{enumerate} 
\end{claim}

\begin{proof} 
If each $\bfu_i$ can be retrieved from a single strand $\bfx_j$ (i.e., $\cC$ is a systematic code), then similarly to the proof of  \autoref{lem:random:single:no code} we have that $T_{\max}^\cC = T_{\text{avg}}^\cC = \E[\tau_i(\cC)] = n$, for any $i\in[n]$. Otherwise, assume w.l.o.g. that $\bfu_1$ cannot be retrieved from a single strand and let $J\subseteq [n]$ be a set of minimal size $|J|=\rho_1$ such that $J \in \cD(1)$. 
By the latter observation and since it is possible to retrieve any information strand $\bfu_i$ from all the $n$ strands, the fact that $n=k$ implies that if there exists $J' \subseteq [n]$, such that $J'$ is a retrieval set of $\bfu_1$ (i.e., $J' \in \widehat{\cD} (1)$) then $J'$ contains $J$.  Hence, $|\cD(1)|=1$, i.e., the set $J$ is the only  minimal retrieval set of $\bfu_1$, and by 
\autoref{claim:random:single:expected maximum retrieval time of encoded strands' subset}, 
we have that 
$ 
\E[\tau_i(\cC)] = \E\left[\widehat{\tau}_{J}(\cC)\right] = n H_{\rho_1} > n$, 
where the last inequality holds since $|J|=\rho_1>1$.
Thus,
$$
T_{\max}^\cC = \max_{1\le i\le k} \E[\tau_i(\cC)] = \max_{1\le i\le n}  n H_{\rho_i} = nH_\rho,
$$
and 
$$
T_{\text{avg}}^\cC =  \frac{1}{k}\sum_ {i=1}^{k} \mathbb{E}[\tau_i(\cC)] = \frac{1}{n}\sum_{i=1}^n nH_{\rho_i} = \sum_{i=1}^n H_{\rho_i}.
$$
Note that $H_{\rho_i}\ge1$ for any $i\in[k]$ and since $\rho_1>1$, we have that $H_{\rho_1}>1$. Hence $T_{\text{max}}^\cC> n$ and $T_{\text{avg}}^\cC> n$ which completes the proof. 
\end{proof} 

\subsection{The Singleton Coverage Depth Problem}

We continue by studying cases where $n>k$. Next, the case where the minimal retrieval sets are disjoint is considered. We start with the case in which a strand $\bfx_i$ has exactly two minimal retrieval sets $\cD(i)=\{A,B\}$ and $A\cap B=\emptyset$, while the next example considers the simple parity code which is a special instance of this case.
\begin{example}\label{ex: random access: parity (4,3)}
    Let $\cC$ be the $(4,3)$ parity code. We have that  $\bfX=(\bfu_1, \bfu_2, \bfu_3, \bfx_4)$, where $${\bfx_4 = \bfu_1 + \bfu_2 + \bfu_3}.$$ Since the code is symmetric, let us consider w.l.o.g. $\bfu_1$. Note that $\cD(1) = \{\{\bfu_1\}, \{\bfu_2, \bfu_3, \bfx_4\}\}$ and the two retrieval sets are disjoint. Hence, we cannot recover $\bfu_1$ from a series of $r$ draws only if the series of draws does not contain $\bfu_1$, and it either contains one unique strand or two unique strands. Hence, 
    \begin{align*}
    \E\left[\tau_1(\cC)\right] & = \sum_{r=0}^\infty P[\tau_i(\cC)>r] = 1 + \sum_{r=1}^\infty \left(3\cdot \frac{1}{4^r} + \binom{3}{2}\sum_{j=1}^{r-1}\binom{r}{j} \frac{1}{4^j}\cdot \frac{1}{4^{r-j}}\right) \\
    & {=1 + 3\sum_{r=1}^\infty \frac{1}{4^r} + 3\sum_{r=1}^\infty \frac{2^r-2}{4^r} =  1 + 3\cdot \frac{1}{3} + 3\cdot \frac{1}{3} = 3.}
    \end{align*}
    That is, in this case, $\E\left[\tau_1(\cC)\right] = k$. 
\end{example}
The next theorem extends \autoref{ex: random access: parity (4,3)} to any code $\cC$ and an information strand $\bfx_i$ with exactly two minimal retrieval sets $A,B$ such that $A\cap B=\emptyset$.

\begin{theorem}
\label{theorem:random:single:two disjoint retrieval sets}
Let $\cC$ be an $(n,k)$ code and $i\in[k]$. If $\cD(i)=\{A,B\}$, for two disjoint retrieval sets, $A\cap B=\emptyset$, then
$\E \left[\tau_i(\cC) \right] = n\cdot \left(H_{|A|} + H_{|B|} - H_{|A|+|B|}\right).$
\end{theorem}

\begin{proof}
Denote $\rho_A = |A|, \rho_B=|B|$. For a set of indices $J \subseteq [n]$, let $\lambda_{J}(r-1)$ be the number of different options to draw strands in the first $r-1$ draws such that for at least one of the indices $j\in J$, the  strand $\bfx_j$ was not drawn. Additionally, let $\lambda(r-1)$ be the number of different options to draw strands in the first $r-1$ draws such that the $i$-th information strand cannot be retrieved from the set of drawn strands.  Note that since $\cD(i)=\{A,B\}$, we have that $\lambda(r-1)$ is the number of different options to draw strands in the first $r-1$ draws such that at least one strand from $A$ and at least one strand from $B$ were not drawn. Hence.
$$ \lambda_{A\cup B}(r-1)  =  \lambda_{A}(r-1) + \lambda_{B}(r-1) - \lambda(r-1),$$
and 
\begin{align*}
\lambda(r-1) & = \lambda_{A}(r-1) + \lambda_{B}(r-1) - \lambda_{A\cup B}(r-1)\\
& \stackrel{(a)}{=} \sum_{j=1}^{\rho_A} \binom{\rho_A}{j} (-1)^{j+1} (n-j)^{r-1} \\ 
& \ \ \ + \sum_{j=1}^{\rho_B} \binom{\rho_B}{j} (-1)^{j+1} (n-j)^{r-1}\\
& \ \ \ - \sum_{j=1}^{\rho_A + \rho_B} \binom{\rho_A + \rho_B}{j} (-1)^{j+1} (n-j)^{r-1},
\end{align*}
{where (a) follows from the inclusion-exclusion principle.} 
Using the tail sum formula for the expectation, we have that
\begin{align*}
\E \left[\tau_i(\cC) \right] & = \sum_{r=1}^\infty \frac{\lambda(r-1) }{n^{r-1}} = \sum_{r=1}^\infty  \sum_{j=1}^{\rho_A} \frac{\binom{\rho_A}{j} (-1)^{j+1} (n-j)^{r-1} }{n^{r-1}} \\ 
& \ \ + \sum_{r=1}^\infty \sum_{j=1}^{\rho_B}\frac{ \binom{\rho_B}{j} (-1)^{j+1} (n-j)^{r-1} }{n^{r-1}}\\ 
& \ \ - \sum_{r=1}^\infty \sum_{j=1}^{\rho_A + \rho_B} \frac{\binom{\rho_A + \rho_B}{j} (-1)^{j+1} (n-j)^{r-1} }{n^{r-1}}.
\end{align*}

Next, we analyze the first term in the latter expression and the other two terms can be analyzed similarly. 
\begin{align*}
& \sum_{r=1}^\infty  \sum_{j=1}^{\rho_A} \frac{\binom{\rho_A}{j} (-1)^{j+1} (n-j)^{r-1} }{n^{r-1}} \\ =
& {\sum_{r=1}^{\infty} \sum_{j=1}^{\rho_A} \binom{\rho_A}{j} (-1)^{j+1} \left( 1 - \frac{j}{n} \right)^{r-1}} \\ \stackrel{(a)}{=}
&   \sum_{j=1}^{\rho_A}  \binom{\rho_A}{j} (-1)^{j+1}\sum_{r=1}^\infty \left(1-\frac{j}{n}\right)^{r-1} \\ 
\stackrel{(b)}{=} & \sum_{j=1}^{\rho_A}  \binom{\rho_A}{j} (-1)^{j+1} \frac{n}{j} = n \sum_{j=1}^{\rho_A}  \binom{\rho_A}{j} \frac{(-1)^{j+1} }{j}  \\ \stackrel{(c)}{=} 
& nH_{\rho_A}.
\end{align*}
We note that (a) holds since the sum is absolutely convergent. $(b)$ follows since $\left\{\left(1-\frac{j}{n}\right)^{r-1}\right\}_{r=1}^\infty$ is a geometric series. The equality $(c)$ can be observed by considering Euler’s integral representation of the harmonic numbers~\cite{Eulerbook}, $H_{\rho_A}=\int_0^1\frac{1-x^{\rho_A}}{1-x}dx$. using the latter we have that 
\begin{align*}
  H_{\rho_A } & = 
\int_0^1\frac{1-x^{\rho_A}}{1-x}dx = \int_0^1 \frac{1-(1-y)^{\rho_A}}{y}dy \\
& = \sum_{j=1}^{\rho_A} \left(\binom{\rho_A}{j}(-1)^{j+1}
 \int_0^1y^{j-1}dy \right)=\sum_{j=1}^{\rho_A}  \binom{\rho_A}{j} \frac{(-1)^{j+1} }{j}.
\end{align*}
Thus, 
$$\E \left[\tau_i(\cC) \right] = n\cdot \left(H_{\rho_A} + H_{\rho_B} - H_{(\rho_A+\rho_B)}\right),$$
which concludes the proof. 
\end{proof}

A direct corollary from \autoref{theorem:random:single:two disjoint retrieval sets} is that \autoref{ex: random access: parity (4,3)} can be generalized to any $(n=k+1,k)$ simple parity code $\cC$, and for any $i\in [k]$ the expected number of draws to retrieve $\bfu_i$ using $\cC$ is exactly $k$. 

\begin{corollary} \label{corollary:simple-parity}
Assume $\cC$ is the $(n=k+1,k)$ simple parity code (i.e., $\bfX=(\bfu_1, \ldots, \bfu_k,\sum_{j=1}^{k}\bfu_j)$). Then, for any $i\in[k]$, we have that, $ \E \left[\tau_i(\cC) \right] = k$ and $T_{\max}^\cC =T_{\text{avg}}^\cC= k$. 
\end{corollary}

The proof of \autoref{theorem:random:single:two disjoint retrieval sets} relies on the inclusion-exclusion principle and can be extended to more than two retrieval sets. Since the proof is technical and repeats the same ideas as the ones from \autoref{theorem:random:single:two disjoint retrieval sets}, it is omitted from the paper. 
\begin{corollary}
\label{corollary:random:single: disjoint retrieval sets} 
Let $\cC$ be an $(n,k)$ code and $i\in[k]$.
If $\cD(i)=\{A_1, A_2,\ldots,A_v\}$ for mutually disjoint retrieval sets, then
$$\E \left[\tau_i(\cC) \right] = n\cdot
\left( 
\sum_{s=1}^v (-1)^{s+1} \hspace{-4ex}\sum_{1\le j_1<\cdots < j_{s}\le v} \hspace{-4ex}H_{(|A_{j_1}| + \cdots +|A_{j_{s}}|)} \right).
$$
\end{corollary}
 


\autoref{corollary:simple-parity} states that the simple parity code does not improve the value of $T_{\max}^\cC$. This observation raises the problem of finding codes that indeed improve this parameter, and next we consider MDS codes for this purpose.  First, recall that by \autoref{lem:random:single:no code}, if no code is used, then we have that $T_{\max} = T_{\text{avg}}  = \E \left[\tau_i\right] = k$ for any $i \in [k]$. On the other hand, assume $\cC$ is a \emph{$k$-non systematic MDS code} 
in which the minimal size of a retrieval set, for each of the information strands is $k$. In other words, any set of less than $k$ encoded strands is not a retrieval set. If $\cC$ is used, then in order to retrieve any specific information strand, one should sample a subset of $k$ distinct encoded strands. Hence, by \autoref{th:MDS}, for any $i \in [k]$, we have that,  $T_{\max}^\cC  = \E \left[\tau_i(\cC) \right] =\sum_{j=0}^{k-1} \frac{n}{n-j} \approx n \log(\frac{n}{n-k})$, while if $\frac{k}{n}=R$ is a constant, we have that  $n \log(\frac{n}{n-k}) = \frac{k}{R}\log(\frac{1}{1-R}) > k$.
The next theorem discusses the case where $\cC$ is a  systematic MDS code and shows that for any such code the expected sample size is exactly $k$. The proof can be found in Appendix~\ref{Appendix: random access example}. 

\begin{theorem} \label{th:MDS_random_access}
Let $\cC$ be a systematic $[n,k]$ MDS code. For any $i \in [k]$ we have that $ \E[\tau_i(\cC)]=k$ and hence $T_{\max}^\cC = T_{\text{avg}}^\cC = k$.
\end{theorem}


\subsection{Reducing the Singleton Coverage Depth Below $k$}
In all the codes we studied so far, the expected number of reads to retrieve a single information strand $\bfu_i$, was at least $k$, which means that these codes do not improve upon the case where no coding is used. Next, we present families of $(n,k)$ codes for which $T_{\max}^\cC < k$. We start with the following example of an $(8,4)$ code.  

\begin{example}\label{ex: random access: expect < k=4}
Let $\cC_{(8,4)}$ be the $(8,4)$ code defined as follows. Let $\bfU_{(8,4)}= ( \bfu_1,\bfu_2,\bfu_3,\bfu_4) \in (\Sigma^\ell)^4$ and let 
$$
\bfX_{(8,4)} = ( \bfu_1,\bfu_2,\bfu_3,\bfu_4, \bfu_1+\bfu_2, \bfu_2+\bfu_3, \bfu_3+\bfu_4, \bfu_4+\bfu_1  )\in (\Sigma^\ell)^8.
$$
Denote $\bfx_{i,j} \triangleq \bfu_i + \bfu_j$ and w.l.o.g. assume that we are interested in retrieving $\bfu_1$. It can be verified that 
$$
\cD(1) = \left\{
\begin{matrix}
\left\{\bfu_1 \right\}, \ \   \left\{\bfu_2, \bfx_{1,2}\right\}, \ \ \left\{\bfu_4, \bfx_{1,4}\right\}, \\
\left\{\bfu_3, \bfx_{2,3}, \bfx_{1,2}\right\}, \ \  \left\{\bfu_3, \bfx_{3,4}, \bfx_{1,4}\right\},   \\
 \left\{\bfu_4, \bfx_{3,4}, \bfx_{2,3}, \bfx_{1,2}\right\}, \ \ \left\{\bfu_2, \bfx_{3,4}, \bfx_{2,3}, \bfx_{1,4}\right\} 
\end{matrix}
\right\}, 
$$
while $\cD(1)$ is given with a slight abuse of notation, in which the retrieval sets are given in terms of the encoded strands rather than their indices to simplify the example.
Let $\cE_{r-1}$ be the random variable that represents 
the number of unique strands that were sampled in the first $r-1$ draws. Since any set of $6$ or more unique strands is a retrieval set of $\bfu_1$,
we have that
\begin{align*}
P\left[\tau_1(\cC_{(8,4)})\ge r\right] & = \sum_{i=1}^5 P\left[\tau_1(\cC_{(8,4)})\ge r | \cE_{r-1}=i\right]\cdot P\left[\cE_{r-1} = i\right]\\
& \ \ + P\left[\tau_1(\cC_{(8,4)})\ge r | \cE_{r-1}\ge 6\right]\cdot P\left[\cE_{r-1} \ge 6\right]\\
& = \sum_{i=1}^5 P\left[\tau_1(\cC_{(8,4)})\ge r | \cE_{r-1}=i\right]\cdot P\left[\cE_{r-1} = i\right].
\end{align*} 

It can be readily verified that $P\left[\tau_1(\cC_{(8,4)})\ge r | \cE_{r-1}=1\right] = \frac{7}{8}$. In case $\cE_{r-1}=2$, there are {$\binom{8}{2}=28$} different pairs of strands, and since $\tau_1(\cC_{(8,4)})\ge r$, we should consider only the pairs from which $\bfu_1$ cannot be retrieved. Note that two of the pairs are in $\cD(1)$ and $7$ additional pairs contain $\bfu_1$. 
Hence we have that $P\left[\tau_1(\cC_{(8,4)})\ge r | \cE_{r-1}=2\right] = \frac{28-9}{28}=\frac{19}{28}$.
Similarly,
there are $\binom{8}{3}=56$ different triples, from which $\binom{7}{2}=21$ contain $\bfu_1$, five more triples contain $\{\bfu_2,\bfx_{1,2}\}$ and do not contain $\bfu_1$, additional five triples contain $\{\bfu_3,\bfx_{1,3}\}$  (and do not contain $\bfu_1$), and two more triples are in $\cD(1)$. That is, 
 $P\left[\tau_1(\cC_{(8,4)})\ge r | \cE_{r-1}=3\right] = \frac{56-21-5-5-2}{56} = \frac{23}{56}$. Using similar counting techniques, it can be shown that  
 $P\left[\tau_1(\cC_{(8,4)})\ge r | \cE_{r-1}=4\right] = \frac{8}{70}$, and  ${P\left[\tau_1(\cC_{(8,4)})\ge r | \cE_{r-1}=5\right] = \frac{1}{56}}$.
Furthermore, using the inclusion-exclusion principle,  it can be proved that 
$$
{P[\cE_{r-1}=i] =} \frac{\binom{8}{i}}{8^{r-1}} \sum_{j=0}^{i-1} \binom{i}{j}(-1)^{j} (i-j)^{r-1}.
$$
 By combining all of the above we obtain that
 \[
 \E[\tau_1(\cC_{(8,4)})] = \sum_{r=1}^\infty P\left[\tau_1(\cC_{(8,4)})\ge r\right] = \frac{403}{105}\approx 3.838 = 0.9595  k .
 \]
\end{example}

\autoref{ex: random access: expect < k=4}
can be extended to any integer $k\ge 2$ as follows. 

\begin{construction} \label{construction_2k_k}
    Let $\cC_{(2k,k)}$ be the $(n=2k,k)$ code such that    
    $$\bfU_{(2k,k)}=(\bfu_1,\bfu_2,\ldots,\bfu_k)\in(\Sigma^\ell)^k$$ and  
    $$\bfX_{(2k,k)} = (\bfu_1,\ldots,\bfu_k, \bfu_1+\bfu_2,\ldots, \bfu_{k-1}+\bfu_k, \bfu_k+\bfu_1)\in(\Sigma^\ell)^{2k}.$$
\end{construction}

Similarly to \autoref{ex: random access: expect < k=4}, the value $\E[\tau_1(\cC_{(2k,k)})]$ can be expressed using the   conditional probabilities $P[\tau_1(\cC_{(2k,k)})\ge r | \cE_{r-1}=i]$. The evaluation of these conditional probabilities can be done using a recursive formula which is given in the next theorem together with the expected value of $\tau_1(\cC_{(2k,k)})$, while the proof appears in Appendix~\ref{Appendix: random access example}. 

\begin{theorem}\label{th: random access: (2k,k) code}
For any $k\ge 2$, and any $j\in [k]$ we have that 
\[
\E[\tau_j(\cC_{(2k,k)})] = 1 + \sum_{i=1}^{2k-3} B(k,i) \cdot \frac{2k}{(2k-i)\binom{2k}{i}},
\]
where 
\[
B(k,i) = \begin{cases}
     \binom{2k-1}{i} + 2B(k-1,i-1) - B(k-2,i-2) & k\ge 2, i\ge 2\\
     1 & k\ge 0, i=0\\
     2k+1 & k\ge 0, i=1\\
     1 & k=1, i=2\\
     0 & k=0, i\ge 2\\
     0 & k=1, i\ge 3
     \end{cases}.
\]
\end{theorem}

Even though we did not solve the recursive formula in \autoref{th: random access: (2k,k) code} to obtain an exact value for $\E[\tau_1(\cC_{(2k,k)})]$, we used it to calculate $\E[\tau_1(\cC_{(2k,k)})]$ for values $2\le k\le 100$ and the results can be found in Fig~\ref{fig:comparison}. Based on these results we have the following conjecture.
\begin{conjecture} For any $k\ge 4$ and any $j\in [k]$, we have that $\E[\tau_j(\cC_{(2k,k)})]< k$. Moreover, the ratio $\frac{\E[\tau_j(\cC_{(2k,k)})]}{k}$ decreases with $k$ and $$ \lim_{k\to\infty} \frac{\E[\tau_j(\cC_{(2k,k)})]}{k} < 0.9456.$$
\end{conjecture}

The following definition is used in the next theorem. 
\begin{definition} \label{def:gamma_block_code}
 Given an $(n,k)$ code $\cC$ as defined above and an integer $\gamma \ge 1$, we say that a $(\gamma n, \gamma k)$ code $\cC^\gamma$ is the \emph{$\gamma$-block code of $\cC$} if for an information word $$\bfU = \bfU_1 \circ \bfU_2 \cdots \circ \bfU_\gamma = (\bfu_1, \ldots, \bfu_k) \circ (\bfu_{k+1}, \ldots, \bfu_{2k}) \circ \cdots \circ (\bfu_{(\gamma-1)k+1}, \ldots, \bfu_{\gamma k}),$$  the corresponding codeword $\cX_\gamma$ satisfies, 
$$E_{\cC^\gamma} (\bfU) = \bfX = \bfX_1 \circ \bfX_2 \circ \cdots \circ \bfX_\gamma = E_\cC ( \bfU_1) \circ E_\cC (\bfU_2) \circ \cdots \circ E_\cC (\bfU_\gamma), $$  where $E_\cC$ denotes the encoder of the code $\cC$. 
\end{definition}

In the next theorem, we show that given an $(n,k)$ code $\cC$, one can increase $k$ by using a $\gamma$-block code $\cC^{\gamma}$, without changing the ratio between the expected number of draws to the number of information strands. 
\begin{theorem}\label{th: random access: code concatenation}
    Let $\cC$ be an $(n,k)$ code. For an integer $\gamma \ge 1$, let $\cC^\gamma$ be a $\gamma$-block code of $\cC$. For any $1 \le i \le \gamma k$, it holds that, 
    $\E [\tau_i (\cC^\gamma)] = \gamma \E[\tau_{i'}(\cC)], $
    where $i' \equiv i \pmod k$ and $1 \le i' \le k$.
\end{theorem}
\begin{proof}
For any $r \ge 1$ draws, let us denote by $\varepsilon_i^r$  the random variable that governs the number of strands drawn from $\bfX_s$ (from the $r$ draws), where $s$ is an integer and $u_i \in \bfU_s.$ Then we have that
\begin{align*}
\E [\tau_i (\cC^\gamma)] & = \sum_{r=1}^\infty P[ \tau_i(\cC^\gamma) \ge r] \\ 
& 
= \sum_{r=1}^\infty \sum_{z=0}^{\infty} P [\varepsilon_i^{r-1} = z]\cdot P\left[\tau_i(\cC^\gamma) \ge r | \varepsilon_i^{r-1} = z\right]
\\
& \stackrel{(a)}{=} \sum_{r=1}^\infty \sum_{z=0}^{r-1} P [\varepsilon_i^{r-1} = z]\cdot P\left[\tau_i(\cC^\gamma) \ge r | \varepsilon_i^{r-1} = z\right] \\ 
& = \sum_{r=1}^\infty \sum_{z=0}^{r-1} \binom{r-1}{z} \left(\frac{1}{\gamma}\right)^z \left(1-\frac{1}{\gamma}\right)^{r-z-1} \cdot P\left[\tau_i(\cC^\gamma) \ge r | \varepsilon_i^{r-1} = z\right]\\
& =   \sum_{r=1}^\infty \sum_{z=0}^{r-1} \binom{r-1}{z} \left(\frac{1}{\gamma}\right)^z \left(1-\frac{1}{\gamma}\right)^{r-z-1} \cdot P \left[ \tau_i(\cC) \ge z+1\right] \\
& = \sum_{z=0}^\infty P \left[ \tau_i (\cC ) \ge z+1 \right] \sum_{r=z+1}^{\infty} \binom{r-1}{z} \left(\frac{1}{\gamma}\right)^z \left(1-\frac{1}{\gamma}\right)^{r-z-1}  \\ 
& = \sum_{z=0}^\infty P \left[ \tau_i (\cC ) \ge z+1 \right] \sum_{r=z}^{\infty} \binom{r}{z} \left(\frac{1}{\gamma}\right)^z \left(1-\frac{1}{\gamma}\right)^{r-z} \\ 
& \stackrel{(b)}{=} \sum_{z=0}^\infty P \left[ \tau_i (\cC ) \ge z+1 \right] \cdot \gamma  \\ 
& = \sum_{z=1}^{\infty} P \left[ \tau_i (\cC) \ge z \right] \cdot \gamma = \gamma \E \left[ \tau_i (\cC) \right],
\end{align*}
where equality (a) follows from the fact that the probability to collect $z>r-1$ unique strands from $\bfX_s$, using only $r-1$ draws is zero for any integer $s$, i.e., $P[\varepsilon_i^{r-1}=z]=0$.  
To see that equality (b) holds, recall that $\sum_{r=0}^\infty x^r=\frac{1}{1-x}$, and by taking the derivative of the latter $z$ times we get $$\sum_{r=z}^\infty r\cdot(r-1)\cdots(r-z+1)x^{r-z}=\frac{z!}{(1-x)^{z+1}},$$ which is equivalent to 
\begin{align*}\label{eq: binom sums}
    \sum_{r=z}^\infty \binom{r}{z}x^{r-z} = \frac{1}{(1-x)^{z+1}}.
\end{align*}
Lastly, by substituting $x=1-\frac{1}{\gamma}$, equality (a) follows.   
\end{proof}

\autoref{th: random access: code concatenation} implies that given an $(n,k)$ code $\cC$, that achieves good results in terms of minimizing the  expressions $\frac{\E [\tau_i (\cC)]}{k}$, for $i \in [k]$, it is possible to construct an infinte family of fixed-rate codes  $\{ \cC ^\gamma \}_{\gamma = 1}^{\infty}$, such that for any integer $\gamma \ge 1 $, $\cC^\gamma$ is an $(\gamma n, \gamma k )$ code and for any $i_\gamma \in [\gamma k] $ there exists $i \in [k] $, such that  $$\frac{\E [\tau_{i_\gamma} (\cC^\gamma)]}{\gamma k} = \frac{\E [\tau_{i} (\cC)]}{k}.$$
That is, for any integer $\gamma \ge 1,$ the code $\cC^\gamma$ has the  \emph{same behavior} as the code $\cC$ in terms of minimizing the normalized expected singleton coverage depth. Hence, combining \autoref{ex: random access: expect < k=4} and \autoref{th: random access: code concatenation} leads to the following corollary.

\begin{corollary} \label{corollary:random:single:block-code-8-4}
For any integer $\gamma \ge 1,$ let $\cC_{(8 \gamma, 4 \gamma)}^{\gamma}$ be the $\gamma$-block code of $\cC=\cC_{(8,4)}$ (see \autoref{ex: random access: expect < k=4}). For any $i_\gamma \in [\gamma k]$, where $k=4$, we have that  
$$\E \left[\tau_{i_\gamma} (\cC_{(8\gamma, 4\gamma)}^\gamma)\right] = T_{\text{max}}^{\cC_{(8\gamma, 4\gamma)}^\gamma} = T_{\text{avg}}^{\cC_{(8\gamma, 4\gamma)}^\gamma} =  0.9595 \gamma k.$$
\end{corollary}

Note that our numerical computations of the expression in \autoref{th: random access: (2k,k) code} imply that the value $\frac{\E[\tau_i(\cC_{(2k,k)})]}{k}$ decreases with $k$, for $2\le k \le 100$. In particular, for  $k>3$, $\frac{\E[\tau_i(\cC_{(2k,k)})]}{k} \le 0.9456$ and thus by \autoref{th: random access: code concatenation} it is possible to construct codes that improve upon the result in \autoref{corollary:random:single:block-code-8-4} for infinite values of $k$. 


Next, we demonstrate that the value $\frac{\E[\tau_i(\cC)]}{k}$ can be further reduced, by letting the rates of our codes vanish.

\begin{construction} \label{construction_ryan}
    Let $n$ be an integer, $p\in(0,1)$, and assume for simplicity that $np$ is an integer that is dividable by $k$. Additionally, let $\cC^{\text{MDS}}_{n,k,p}$ be a $[n(1-p)+k, k]$ systematic MDS code. We define the $(n,k)$ code $\cC^k_{n,p}$ as follows. For $\bfU=(\bfu_1,\bfu_2, \ldots, \bfu_k)\in(\Sigma^\ell)^k$, let $$ (\bfu_1,\bfu_2, \ldots, \bfu_k, \bfx_1, \bfx_2, \ldots, \bfx_{(1-p)n})\in(\Sigma^\ell)^{n(1-p)+k}$$ be the encoding of $\bfU$ using the encoder of $\cC^{\text{MDS}}_{n,k,p}$. Then,  
    \[
    \bfX = (\underbrace{\bfu_1,\ldots, \bfu_1}_{\frac{pn}{k} \text{ times}},\ \underbrace{\bfu_2,\ldots, \bfu_2}_{\frac{pn}{k} \text{ times}},\ldots, \underbrace{\bfu_k,\ldots, \bfu_k}_{\frac{pn}{k} \text{ times}}, \bfx_1, \bfx_2, \ldots, \bfx_{(1-p)n})\in(\Sigma^\ell)^n.
    \]    
\end{construction}

\begin{theorem}\label{th: random access: zero rate construction}
    For $k=2,3$, there exists $p_2, p_3 \in (0,1)$, such that for any information strand $i\in [k]$, we have that,  $$ \E [\tau_i (\cC^2_{n,p_2})] \approx 1.83 = 0.9143k, $$ and 
    $$ \E [\tau_i (\cC^3_{n,p_3})] \approx 2.67 = 0.89k. $$
\end{theorem}

\begin{proof}
We prove the claim only for $k=2$, while the proof for $k=3$ relies on the exact same ideas. Assume w.l.o.g. that we want to retrieve $\bfu_1$. For simplicity of the analysis, also assume that $\bfX$ contains two information stands $\bfu_1,\bfu_2$ (without multiplicity), and that each of them can be drawn with probability $\frac{p}{2}$.
    First note that since $(\bfu_1,\bfu_2, \bfx_1, \bfx_2, \ldots, \bfx_{(1-p)n})$ belongs to a $[(1-p)n+2,2]$ MDS code, any two distinct strands form a retrieval set for $\bfu_1$, and hence the only case in which we didn't retrieve $\bfu_1$ in $r$ draws, is when we draw the same strand (which is not $\bfu_1$) $r$ times.
    \begin{itemize}
        \item $\tau_1 (\cC^2_{n,p})=1$ only in case we draw $\bfu_1$ in the first draw which happens with probability $\frac{p}{2}$.
        \item $\tau_1 (\cC^2_{n,p})=r$ for $r\ge 2$ only if the first $r-1$ draws are of the strand $\bfx\ne\bfu_1$ and the last draw is of a different strand. Hence we have that
        \[
        P[\tau_1 (\cC^2_{n,p})=r] = \left(\frac{p}{2}\right)^{r-1}\left(1-\frac{p}{2}\right) + (1-p)n\cdot \left(\frac{1}{n}\right)^{r-1}\left(1-\frac{1}{n}\right).
        \]
    \end{itemize}
    Thus, 
    \begin{align*}
    \E [\tau_1 (\cC^2_{n,p})] & = \sum_{r=1}^\infty P[ \tau_i(\cC_{n,p}) = r] \cdot r
    \\ & = \frac{p}{2} + \sum_{r=2}^\infty r \left(\left(\frac{p}{2}\right)^{r-1}\left(1-\frac{p}{2}\right) + (1-p)n\cdot \left(\frac{1}{n}\right)^{r-1}\left(1-\frac{1}{n}\right)\right)\\
    & = \frac{p}{2} + \left(1-\frac{p}{2}\right) \sum_{r=2}^\infty r \left(\frac{p}{2}\right)^{r-1}  + (1-p)\left(1-\frac{1}{n}\right)\sum_{r=2}^\infty r\left(\frac{1}{n}\right)^{r-2}.
\end{align*}
For $n$ large enough $(1-\frac{1}{n})\sum_{r=2}^\infty r\left(\frac{1}{n}\right)^{r-2}\approx 2$ and hence for $n$ large enough we have that 
    \begin{align*}
    \E [\tau_1 (\cC^2_{n,p})] & \approx
     \frac{p}{2} + \left(1-\frac{p}{2}\right) \sum_{r=2}^\infty r \left(\frac{p}{2}\right)^{r-1}  + 2(1-p)\\
     & = \frac{p}{2} +  \left(1-\frac{p}{2}\right) \frac{p(4-p)}{(2-p)^2} + 2(1-p) \\
     & = \frac{p}{2} + \frac{p(4-p)}{2(2-p)} + 2(1-p).
\end{align*}
This expression is minimized when $p=2-\sqrt{2}$ and in this case we have
$$
\frac{p}{2} + \frac{p(4-p)}{2(2-p)} + 2(1-p) \approx 1.83.
$$
Note that even though the optimal $p$ is irrational, since the latter function is continuous for $p\in(0,1)$, we can get as close as we want to this optimum value. Hence, we have that 
$$ \E [\tau_1 (\cC^2_{n,p})] \approx 1.83 = 0.9143k.$$ 
\end{proof}

Combining \autoref{th: random access: code concatenation} and \autoref{th: random access: zero rate construction} leads to the following corollary.

\begin{corollary} 
Let $p_2,p_3\in(0,1)$ be the constants from \autoref{th: random access: zero rate construction}.
For any integer $\gamma \ge 1,$ let $\cC_{n, p_2}^{2,\gamma}$ be the $(n\gamma, 2\gamma)$ $\gamma$-block code of $\cC=\cC_{n, p_2}^{2}$, and similarly let $\cC_{n, p_3}^{3,\gamma}$ be the $(n\gamma, 3 \gamma)$ $\gamma$-block code of $\cC= \cC_{n, p_3}^{3}$  (see \autoref{def:gamma_block_code}). For any $i_2 \in [2\gamma]$, and $i_3\in[3\gamma]$ we have that  
$$\E [\tau_{i_2} (\cC^{2,\gamma}_{n,p_2})] = T_{\text{max}}^{\cC^{2,\gamma}_{n,p_2}} = T_{\text{avg}}^{\cC^{2,\gamma}_{n,p_2}} \approx 1.83\gamma = 0.9143\cdot (2\gamma),$$
and 
$$
\E [\tau_{i_3} (\cC^{3,\gamma}_{n,p_3})] = T_{\text{max}}^{\cC^{3,\gamma}_{n,p_3}} = T_{\text{avg}}^{\cC^{3,\gamma}_{n,p_3}} \approx 2.67\gamma = 0.89\cdot (3\gamma).
$$
\end{corollary}

The evaluation of $\E [\tau_i (\cC^k_{n,p})]$ for $k>3$ can be done using the same technique, however, it becomes less elegant and we do not attempt to evaluate the latter expression rigorously. Nevertheless, we did try to gain a better understanding of the behavior of these codes by computer simulations as follows. Each simulation was done while fixing $k \in \{1, \ldots ,10\} \cup \{ 20, 30, \ldots 100\}$ and $p \in \{ 0.2, 0.4, 0.6, 0.8\}$. For each pair of $k$ and $p$ we started by encoding $k$ information strands with an $[\lfloor(1-p)n\rfloor+k, k]$ MDS code $\cC^{\text{MDS}}_{n,k,p}$, from which we constructed the $(n,k)$ code $\cC^k_{n,p}$ (see \autoref{construction_ryan}). Then, we simulated the sampling process by picking a single strand at each draw (with an equal probability of $\frac{1}{n}$). The simulation stops whenever we can recover $\bfu_1$. We repeated this process $10^7$ times for each pair of $k$ and $p$ and plotted the mean number of the required draws, which is an empirical approximation of $\E [\tau_i (\cC^k_{n,p})]$. Our simulations imply that for most of the tested values of $k$, the optimal value of  $p$ is around $0.6$. Furthermore, it can be seen that $\E [\tau_i (\cC^k_{n,p})]$ decreases as $k$ increases. Finally, it should be noted that even though such codes are not applicable, they allow us to gain insights about the achievable values of  $\E [\tau_1 (\cC)]$.

\begin{figure}
    \centering
    \includegraphics[width=.85\linewidth]{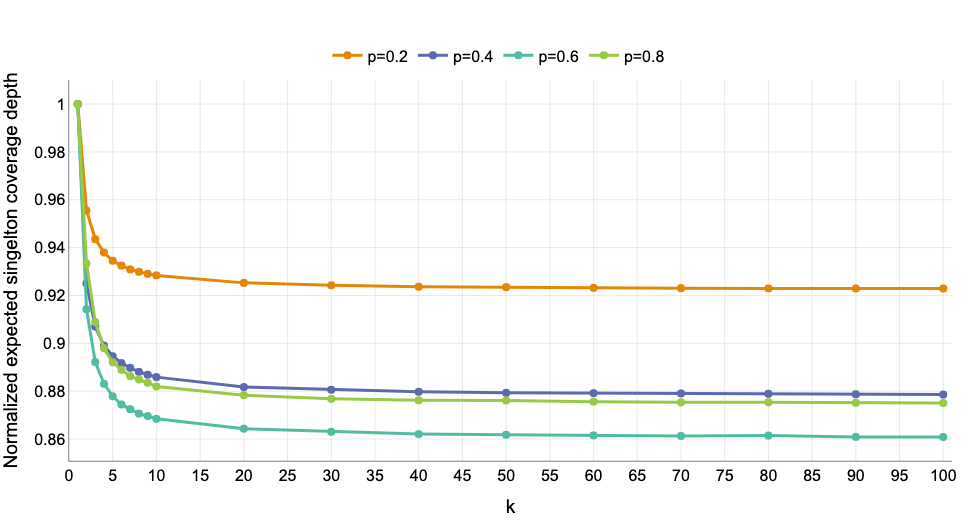}
    \caption{Approximated values of $\E [\tau_i (\cC^k_{n,p})]$ (\autoref{construction_ryan}) for different values of $p \in \{ 0.2, 0.4, 0.6, 0.8 \}$ as a function of $k \in \{1,2,\ldots, 10\} \cup \{ 20, 30, \ldots, 100\}$, where $n=10^8$. The approximated values were obtained empirically by $10,000,000$ computer simulations per any pair of values of $k$ and $p$. The presented results are normalized by $k$.}
    \label{fig:construction ryan}
\end{figure}

\subsection{Lower Bounds}
This section concludes with lower bounds on the value of $\E[\tau_i(\cC)]$. 

\begin{lemma} \label{lm:lowerboundra}  
For any $(n,k)$ code $\cC$, $T^{\cC}_{\max} \geq \frac{k+1}{2}$.
\end{lemma}

\begin{proof}
Assume the word $\bfU$ was encoded to the codeword $\bfX$. Every sequence of reads can be expressed as a vector $\bfv\in[n]^*$, and for every such a $\bfv$, denote by $n_i(\bfv)$, for $i\in[k]$, the minimum read index $h$ which allows retrieving the $i$-th information strand $\bfu_i$. The key intuition behind our approach is that each new sample collected during the sequence of reading the strands allows us to recover at most one new information strand. Hence,
$$\sum_{i=1}^k n_i(\bfv)= n_1(\bfv)+n_2(\bfv)+\cdots +n_k(\bfv)\geq \sum_{i=1}^k i = k(k+1)/2.$$ 
Hence, it follows that, $\sum_{i=1}^k\tau_i(\cC)\geq k(k+1)/2$ and therefore 
$$\E\left[\sum_{i=1}^k \tau_i(\cC)\right]=  \sum_{i=1}^k\E[ \tau_i(\cC)] \geq k(k+1)/2.$$ In particular, there exists $i\in[k]$ for which $\E[ \tau_i(\cC)] \geq \frac{k+1}{2}$, i.e., $T^{\cC}_{\max} \geq \frac{k+1}{2}$.     
\end{proof}

Even though the bound in \autoref{lm:lowerboundra}
holds for any code $\cC$, it appears that in most cases the bound is not tight. To obtain a tighter lower bound on $T_{\max}^\cC$ in the next theorem we also consider the rate of the code $\cC$.

\begin{theorem}\label{th: ramdom access: improved lower bound}
    Let $\cC$ be an $(n,k)$ code. It holds that 
    \[
    T_{\max}^\cC \ge \frac{n}{k}\cdot \sum_{i=0}^k\frac{k-i}{n-i} = n-\frac{n(n-k)}{k}\cdot(H_n-H_{n-k}).
    \]
\end{theorem}
\begin{proof}
Let us use the same notations
as in the proof of \autoref{lm:lowerboundra}. Additionally, denote by $t_i(v)$ the time to collect the $i$-th new sample (after collecting the previous one). Clearly, we have that
\[
\sum_{i=1}^k \tau_i(\cC) = \sum_{i=1}^k n_i(v)\ge \sum_{i=1}^k \sum_{j=1}^i t_j(v).
\]

Define $t_j(\cC)$ to be the random variable that governs the time to collect the $j$-th new sample (after collecting the previous one). Hence, 
    \begin{align*}
        \sum_{i=1}^k \E \left[\tau_i(\cC)\right] = \E\left[\sum_{i=1}^k \tau_i(\cC)\right] \ge \E\left[\sum_{i=1}^k \sum_{j=1}^i t_j(\cC)\right] = \sum_{i=1}^k\sum_{j=1}^i \E\left[t_j(\cC)\right].
    \end{align*}
    Note that for any $j\in[k]$, we have that $t_j(\cC)$ is a geometric random variable with success probability $p_j=\frac{n-(j-1)}{n}$  and so $\E[t_j(\cC)] = \frac{n}{n-(j-1)}$, and
    \begin{align*}
    \sum_{i=1}^k \E \left[\tau_i(\cC)\right] & \ge  \sum_{i=1}^k \sum_{j=1}^i \E\left[t_j(\cC)\right] 
    = \sum_{i=1}^k \sum_{j=1}^i \frac{n}{n-(j-1)}\\
    & = n \sum_{i=1}^k \left(\frac{1}{n-i+1} + \frac{1}{n-i+2} + \ldots + \frac{1}{n}\right) \\
    & = n\left(\frac{k}{n} + \frac{k-1}{n-1} + \ldots + \frac{1}{n-k+1}\right)  = n\sum_{i=0}^{k-1}\frac{k-i}{n-i}.
    \end{align*}
 For any $i\in[k]$, we have that 
    \begin{align*}
        \frac{k-i}{n-i} = \frac{k}{n} - \left(1-\frac{k}{n}\right)\frac{i}{n-i} ,
    \end{align*}
    which implies that
    \begin{align*}
       n \sum_{i=0}^k\frac{k-i}{n-i} &=  n\sum_{i=0}^{k-1} \left(\frac{k}{n} - \left(1-\frac{k}{n}\right)\frac{i}{n-i} \right) \\
       & = k^2 - n\left(1-\frac{k}{n}\right)\sum_{i=0}^{k-1}\frac{i}{n-i}\\
       & = k^2 - (n-k)\sum_{i=0}^{k-1}\left(\frac{n}{n-i}-1\right)\\
       & = k^2 + k(n-k) - n(n-k)\sum_{i=0}^{k-1}\frac{1}{n-i} \\
       & = nk - n(n-k)(H_n-H_{n-k}).
    \end{align*}
    Hence we have that 
    \begin{align*}
    \frac{1}{k}  \sum_{i=1}^k \E \left[\tau_i(\cC)\right] & \ge \frac{n}{k}\cdot \sum_{i=0}^k\frac{k-i}{n-i} = n - \frac{n(n-k)}{k}\cdot (H_n-H_{n-k}).
    \end{align*}
    In particular, there exists $i\in[k]$ for which $\E[ \tau_i(\cC)]  \ge \frac{n}{k}\cdot \sum_{i=0}^k\frac{k-i}{n-i} = n - \frac{n(n-k)}{k}\cdot (H_n-H_{n-k})$, i.e., $T^{\cC}_{\max}  \ge \frac{n}{k}\cdot \sum_{i=0}^k\frac{k-i}{n-i} = n - \frac{n(n-k)}{k}\cdot (H_n-H_{n-k})$. 
    \end{proof}
    Lastly, we conclude with the following lemma.

    \begin{corollary} \label{corollary:lower bound rate}
        Let $0<R<1$ and consider a sequence of codes $\{\cC_i\}_{i=1}^{\infty}$ with parameters $(n_i,k_i)$ such that for any $i$, $n_i<n_{i+1}$, and $R = \frac{k_i}{n_i}$. It holds that,

        \[
        \lim_{i \to \infty} \frac{T_{\max}^{\cC_i}}{k_i} \ge \left(\frac{1}{R}+\frac{1-R}{R^2}\cdot \log(1-R)\right).
        \]
        That is, for any $\varepsilon>0$, there exists $i$ large enough (i.e., $n_i, k_i$ large enough) such that,
        \[
        {T_{\max}^{\cC_i}} \ge k_i\left(\frac{1}{R}+\frac{1-R}{R^2}\cdot \log(1-R)\right) - \varepsilon.
        \]
    \end{corollary}
    \begin{proof}
        From \autoref{th: ramdom access: improved lower bound}, for any $(n,k)$ code $\cC$, we have that 
        \begin{align*}
            T_{\max}^\cC & \ge n - \frac{n(n-k)}{k}\cdot (H_n-H_{n-k}).
        \end{align*}
Thus, we have that 
        \begin{align*}
        \lim_{i \to \infty } \frac{T_{\max}^{\cC_i}}{k_i}  &\ge \lim_{i \to \infty } \frac{1}{k_i} \left( n_i - \frac{n_i(n_i-k_i)}{k_i}\cdot (H_{n_i}-H_{n_i-k_i}) \right) \\
             & =     \lim_{i \to \infty } \frac{n_i}{k_i} \left( 1 - \frac{n_i-k_i}{k_i}\cdot (H_{n_i}-H_{n_i-k_i}) \right)  \\
             & =    \lim_{i \to \infty }  \frac{1}{R} - \frac{ (1-R)}{R^2}(H_{n_i}-H_{n_i-k_i})
             \\ & =  
             \frac{1}{R} - \frac{1-R}{R^2}\log\left(\frac{1}{1-R}\right). 
        \end{align*}

        It can be verified that in this case if $R$ approaches zero, one, then the latter expression approaches~$\frac{1}{2}, 1$, respectively. 
       
    \end{proof}

\begin{figure}
    \centering
    \includegraphics[width=.85\linewidth]{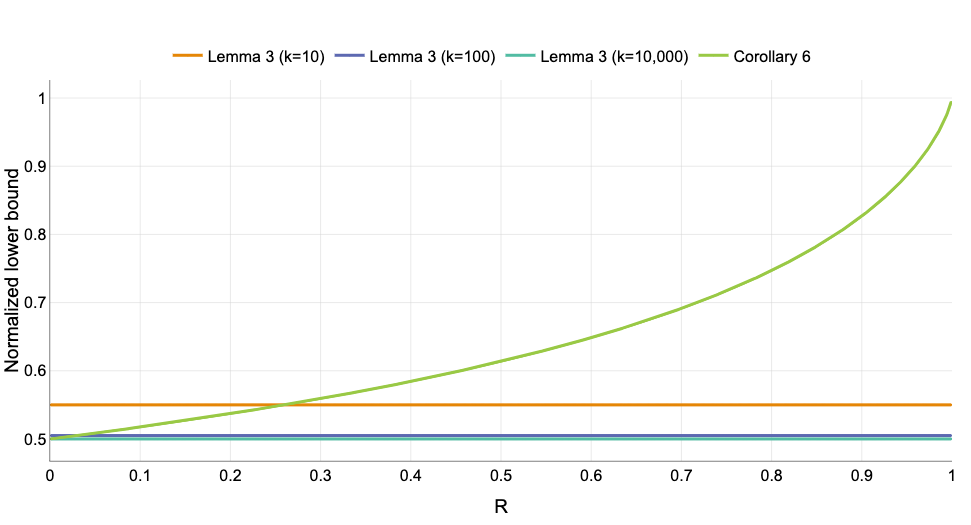}
    \caption{Comparison of the lower bounds (\autoref{lm:lowerboundra} and \autoref{corollary:lower bound rate}) as a function of the rate $R=\frac{k}{n}$. The presented results are normalized by $k$. }
    \label{fig:lower bounds}
\end{figure}

\autoref{fig:lower bounds} presents a comparison between the lower bounds of \autoref{lm:lowerboundra} and \autoref{corollary:lower bound rate} as a function of the code rate $R = \frac{k}{n}$. As can be seen in the figure, in most cases, the bound in \autoref{corollary:lower bound rate} is tighter than the one from \autoref{lm:lowerboundra}. More than that,  the code rate from which the bound in \autoref{corollary:lower bound rate} is tighter than the bound from \autoref{lm:lowerboundra} decreases with $k$.

\begin{figure}
    \centering
    \includegraphics[width=.85\linewidth]{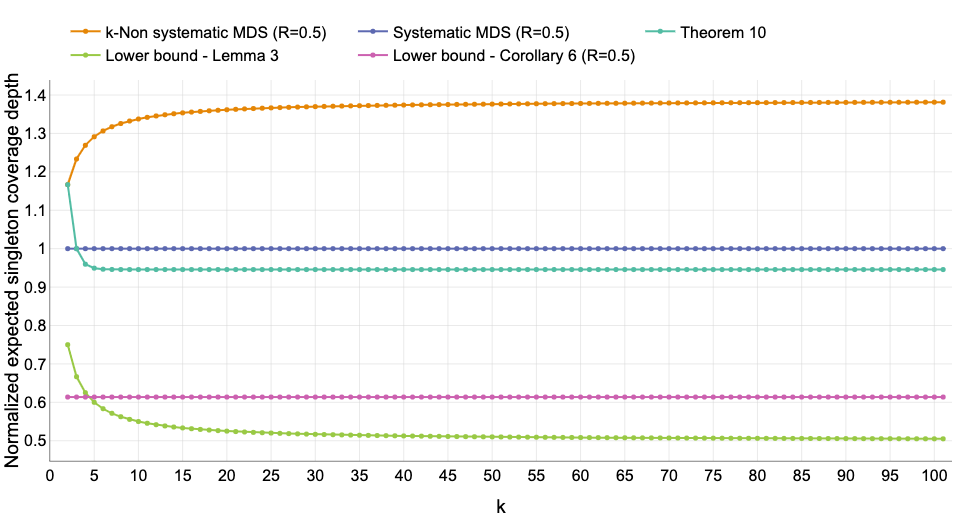}
    \caption{Comparison of the normalized expected singleton coverage depth for code with rate $R=0.5$. }
    \label{fig:comparison}
\end{figure}

Finally, we give in \autoref{fig:comparison} a comparison of the normalized expected singleton coverage depth  for different codes with rate of exactly $R=0.5$. It can be seen in the figure that the $k$-non systematic MDS code achieves the worst results, while the code in \autoref{th: random access: (2k,k) code} achieves the best results, which are roughly $55\%$ lower than the $k$-non systematic MDS code and roughly $10\%$ lower than a systematic MDS code. To offer a better understanding of these results, the lower bounds discussed in \autoref{lm:lowerboundra} and \autoref{corollary:lower bound rate} are also given in the figure.

\section{Conclusion}\label{sec:conc}

In this paper, we have introduced and extensively investigated the novel problem of DNA coverage depth, aiming to reduce sequencing costs and latency while ensuring high-accuracy retrieval.
Our contributions encompass the MDS coverage depth problem, demonstrating the superiority of MDS codes in the noiseless channel. For noisy channels, we proved several bounds on the probability of successfully retrieving the information for a given sample size.  Additionally, we have explored the singleton coverage depth problem, revealing insights into code properties and retrieval times, as well as presenting code constructions that can improve the retrieval time. These findings collectively provide a foundational framework for designing efficient and reliable DNA storage systems, with potential implications for advancing the field.

Nonetheless, future research should address the diverse challenges posed by different noise models, investigate coding schemes beyond MDS codes, and extend the coverage depth problem for additional scenarios. Several possible directions and open problems are listed below. 
\begin{enumerate}
    \item Extend the results presented in this paper with respect to \autoref{pr:expectation_prob} from a uniform distribution to additional channel distributions $\bfp$; e.g. the normal distribution. 
    \item In this work, the noisy channel was modeled by a parameter $t$, under the assumption that retrieval succeeds with probability $1$ given $t$ or more noisy copies and fails otherwise. A very relevant extension to this noise model is to consider the more realistic behavior of the channel, in which the success probability can increase or decrease as a function of the number of noisy copies, i.e., as a function of the cluster size.  
    \item Define and study the coverage depth random access problem for the case in which a subset of size greater than one should be retrieved. This can be considered for arbitrary subsets of the information strands or for pre-defined subsets, that represent units of information (e.g. files). 
    \item Study the coverage depth random access problem under the assumption of noisy channel and/or channel with non-uniform distribution.
\end{enumerate}
\section*{Acknowledgement}

The authors thank John M. Hoffman for raising up the real-world necessity of minimizing the coverage depth and understanding how it can be done with coding and to Zohar Yakhini for helpful discussions about the theoretical definition of the model. The authors also thank Tomer Cohen for analyzing the code presented in \autoref{construction_2k_k}. Lastly, the authors thank Ron M. Roth for suggesting the current version of the proof of \autoref{lem:uni-opt}, which is more elegant than its original version.

\newpage
\appendices

\section{}
\label{Appendix_ExpectationBound_f}

\begin{claim} \label{claim:binomial}
For $n>16$, it holds that, 
\begin{align*}
\sum_{j=0}^{t-1}\binom{r}{j} \left( \frac{1}{n} \right)^{j} \left( 1- \frac{1}{n} \right)^{r-j}  
 {\le t \cdot \binom{r}{t-1} \left( \frac{1}{n} \right)^{t-1} \left( 1-\frac{1}{n} \right)^{r -(t-1)} }
\end{align*}
\end{claim}

\begin{IEEEproof}
To prove the claim, we show that for $0 \le j \le t-1$,  
\begin{align*}
\binom{r}{j} \left( \frac{1}{n} \right)^{j} \left( 1- \frac{1}{n} \right)^{r-j}  \le \binom{r}{j+1} \left( \frac{1}{n} \right)^{j+1} \left( 1- \frac{1}{n} \right)^{r-(j+1)}. 
\end{align*}
The latter can be proved by showing the following equivalent inequality   
$$n \le \frac{r-j}{j+1} \left( 1 - \frac{1}{n} \right)^{-1}. $$

The expression $\frac{r-j}{j+1} \left( 1 - \frac{1}{n} \right)^{-1}$ is minimized at $j=t-1$ (considering only $j\in\{0, 1, \ldots, t-1\}$), thus it is enough to show that, $n \le \frac{r-t+1}{t} \left( 1 - \frac{1}{n} \right)^{-1} = \frac{r-t+1}{t} \left( \frac{n}{n-1} \right), $
which follows if $$r=r(n,k,t)  \ge t(n-1) +(t-1).$$ 
Lastly, it can be verified that $r(n,k,t) \ge t(n-1) +(t-1)$ for any integers $n>16$ and $t>1$. 
\end{IEEEproof}

\subsection*{Proof of \autoref{th:concentrate_fixed}}

Denote $r_f \triangleq r_f(n,k,t)$. Similarly to the proof of \autoref{th:concentrate}, when $n$ is large enough, the probability that urn $i$ has at most $t-1$ balls after $r_f$ draws is denoted by $P(z_i(n,r_f) \le t-1)$, where $z_i(n, r_f)$ is a defined as in the proof of \autoref{th:concentrate}. Thus, the probability  is given by, 
\begin{align*}
P(z_i(n,r_f) \le t -1 )  & \le 3  t \cdot \left( \frac{r_f}{n} \right)^{t-1}  \left( 1-\frac{1}{n} \right)^{n \left( \frac{r_f}{n} -\frac{t-1}{n} \right)} 
\\ & \le 6 t^t \frac{(2 \cdot f(n))^{t-1}}{e^{t \cdot f(n)}} \cdot \left(1-R \right) 
\end{align*}
 
Now let us define a random variable $Y$ as the number of urns with less than $t$ balls. As in the proof of \autoref{th:concentrate}, we have that, 
$$P(E_{t}^{(r_f)}) = P(Y \ge n -k +1) \le 6 t^t \frac{(2 \cdot f(n))^{t-1}}{e^{t \cdot f(n)}} (1-R),$$
where the last inequality follows from Markov's inequality.

\section{}
\label{appendix:proof_of_expectation}
\begin{theorem}\label{th:concentrate_for_exp}
 For any $\varepsilon>0$, $n>e^{\frac{6t\cdot 2^{t-1}}{\varepsilon}} \ge 15$, and any integer $h \ge 1$, we have that,
 
\begin{align*}
P \left [ {\nu_{t}(n,k)}  > h \cdot {r(n,k,t) } \right ] < \varepsilon \cdot \frac{h^{t-1}}{\log^{t(h-1)}(n)}. 
\end{align*}
\end{theorem} 

\begin{proof}
Denote $r \triangleq r(n,k,t)$ and recall that within the context of the urn problem (see \autoref{sec:sequencing coverage:noiseless}), the random variable $\nu_{t}(n,k)$ denotes the number of balls (or rounds) necessary to guarantee that we have a set of $k$ urns where each urn has at least $t$ balls. 

If $r$ balls are drawn, we show that the probability that there are at least $k$ urns each with at least $t$ balls is approaching one when $n$ grows. Analogous to the approach used in the previous {section}, we will show that if the number of balls thrown is at least $r$, then the probability to have at most $n-k+1$ urns which are \emph{not} filled with $t$ balls is approaching zero.

The approach leveraged in this section is inspired by a technique first employed by Erd\H{o}s and R\'{e}nyi in~\cite{ErdosReyni}. 
First, we define the following event.
\begin{description}
\item[$E^{(r)}_{t}$:]  After $r$ rounds, there exists a set $S_{t}$, of $n-k+1$ urns, each containing less than $t$ balls. 
\end{description}
Next, we show that the probability of $E^{(r)}_{t}$ approaches zero when $n$ is large. We first define $z_i(n,r)$ for $1\le i \le n$, as a random variable that governs the number of balls in the $i$-th urn, after $r$ draws. 
For $n$ large enough, the probability that urn $i$ has at most $t-1$ balls after $h \cdot r$ draws is denoted by $P(z_i(n,r) \le t-1)$ 
and is given by, 
\begin{align*}
P(z_i(n,h\cdot r) \le t -1 ) 
&= \sum_{j=0}^{t-1}\binom{h\cdot r}{j} \left( \frac{1}{n} \right)^{j} \left( 1- \frac{1}{n} \right)^{h\cdot r-j}  
\\ & \le t \cdot \binom{h\cdot r}{t-1} \left( \frac{1}{n} \right)^{t-1} \left( 1-\frac{1}{n} \right)^{h\cdot r -(t-1)} 
\\& \le t \cdot \left( \frac{h\cdot r\cdot  e}{t-1} \right)^{t-1} \left( \frac{1}{n} \right)^{t-1} \left( 1-\frac{1}{n} \right)^{h\cdot r -(t-1)},
\end{align*}  
where the last inequality follows from the fact that $\binom{h\cdot r}{t-1} \le \frac{hre}{t-1}^{t-1}$. 
Note that ${(\frac{e}{t-1}})^{t-1} <3$, for $t>1$. Thus, 
\begin{align*}
P(z_i(n,r) \le t -1 )
& \le 3  t \cdot \left( \frac{hr}{n} \right)^{t-1}  \left( 1-\frac{1}{n} \right)^{n \left( \frac{r}{n} -\frac{t-1}{n} \right)}.
\end{align*}  
We have that, 
\begin{align*}
P \left(z_i(n,r)\le t -1 \right) & \le 3 t \cdot \left( h \log \left( \frac{n}{n-k} \right) + h t \log \log (n) +  2 h \log(t+1) \right)^{t-1} \left(e^{\left( \frac{-hr}{n} +\frac{t-1}{n} \right)} \right)
\\ & \le 3 t \cdot (2 h \log n)^{t-1} \left(\frac{n-k}{n} \right)^h \left( \frac{1}{\log^{ht} n}\right) \left( \frac{1}{(t+1)^{2h}}\right) e^{\frac{t-1}{n}}
\\ &= 3t \cdot  \frac{e^{\frac{t-1}{n}}}{(t+1)^{2h}}\cdot   \frac{(2h \log n)^{t-1}}{\log^{ht}(n)} \cdot \left(\frac{n-k}{n} \right)^h 
\\ &= 3t \cdot  \frac{e^{\frac{t-1}{n}}}{(t+1)^{2h}}   \cdot \frac{(2h)^{t-1}}{\log^{t(h-1)+1}(n)} \cdot \left(\frac{n-k}{n} \right)^h , 
\end{align*}
where the second inequality holds since for $n$ large enough  $h\log(\frac{n}{n-k}) + t h\log \log (n) + 2 h\log (t+1) \le (2h \log n)$.
It should be noted that for $n>t$, we have that  $3 t \cdot  \frac{e^{\frac{t-1}{n}}}{(t+1)^{2h}}  \le 6t$, and hence, 

$$ P \left(z_i(n,r)\le t -1 \right) \le 6t \cdot \frac{(2h)^{t-1}}{\log^{t(h-1)+1}(n)} \cdot \left( \frac{n-k}{n} \right)^{h}.$$
Now let us define a random variable $Y$ as the number of urns with less than $t$ balls. From the linearity of expectation, regardless if the urns are independent or not, the expected number of urns that have less than $t$ balls is,
\begin{align*}
\lim_{n \to \infty}\E [Y] & = \lim_{n \to \infty} \Sigma_{i=1}^n \E[z_i(n,r)] 
\\ & = \lim_{n \to \infty} n  P\left( z_i(n,r)\le t-1 \right) \le  \lim_{n \to \infty} (n-k) \left ( \frac{n-k}{n} \right)^{h-1} \cdot 6t \cdot \frac{(2h)^{t-1}}{\log^{t(h-1)+1}(n)} 
\end{align*}
Note that 
$$P(E_{t+1}^{(r)}) = P(Y \ge n -k +1) $$
Using Markov's inequality with $n-k+1$ as the parameter we can conclude that, 
$$P(Y \ge n -k +1 ) \le \left ( \frac{n-k}{n} \right)^{h-1} 6t \cdot \frac{(2h)^{t-1}}{\log^{t(h-1)+1}(n)}  . $$
Let us denote $\varepsilon^* = 6t \cdot \frac{2^{t-1}}{\log(n)}$, then we get that, 
$$P(Y \ge n -k +1 ) \le \varepsilon^* \left ( \frac{n-k}{n} \right)^{h-1} \cdot \frac{(h)^{t-1}}{\log^{t(h-1)}(n)} \le \varepsilon^*  \cdot \frac{(h)^{t-1}}{\log^{t(h-1)}(n)}   . $$
Hence, we get that $P(E_{t}^{(r)}) \to 0$ for $n$ large enough which implies the statement in the theorem. 
\end{proof}

\section{}
\label{Appendix_LowerBound_r}
In this appendix, we prove \autoref{the: lower bound r}. The proof is partially based on   \autoref{claim::condition for E[X]<=n-k} which is proven next.  
\subsection*{Proof of \autoref{claim::condition for E[X]<=n-k}: }
Recall that by (\ref{eq: upper bound on E[X(r)]}), for $r\ge nt$ we have that
\begin{align*}
E [X^{(r)}] \leq n  e^{-(t-1) \log_2\frac{n(t-1)}{r}- (r-(t-1)) \log_2 \frac{(r-(t-1))n}{r(n-1)}}.
\end{align*}
Hence, a sufficient condition for $E [X^{(r)}] \leq n-k = n (1 -R)$, it that
\begin{align}\label{eq: sufficient cond}
e^{-(t-1) \log_2\frac{n(t-1)}{r}- (r-(t-1)) \log_2 \frac{(r-(t-1))n}{r(n-1)}} \leq (1 -R).
\end{align}
Note that
\begin{align*}
e^{-(t-1) \log_2\frac{n(t-1)}{r}- (r-(t-1)) \log_2 \frac{(r-(t-1))n}{r(n-1)}}
 & =  e^{-\frac{t-1}{\ln 2} \ln(\frac{n(t-1)}{r})- \frac{(r-(t-1))}{\ln 2}  \ln \left(\frac{(r-(t-1))n}{r(n-1)}\right) }
\\ & = \left(\frac{n(t-1)}{r}\right)^{-\frac{t-1}{\ln 2}} \left( \frac{(r-(t-1))n}{r(n-1)}\right)^{-\frac{(r-(t-1))}{\ln 2}}
\\ & = \left(\frac{r}{n(t-1)}\right)^{\frac{t-1}{\ln2}} \left(\frac{rn-r}{rn-(t-1)n}\right)^{\frac{r-(t-1)}{\ln2}},
\end{align*}
and by denoting $r=\beta n(t-1)$, for some $\beta\ge 1$ we can rewrite the sufficient condition in (\ref{eq: sufficient cond}) as follows.
\begin{align}\label{eq: sufficient cond2}
 \beta \left(1 - \frac{(\beta-1)}{\beta n-1}\right)^{\beta n -1} \le (1-R )^{\frac{\ln2}{t-1}}.
\end{align}
By the definition of $e$, for any constant $\beta$ we have that ${\left(1 - \frac{(\beta-1)}{\beta n-1}\right)^{\beta n -1} \le e^{-(\beta-1)}}$. Hence, if
$
\beta e^{-\beta} \le \frac{1}{e}(1-R)^{\frac{\ln2}{t-1}}
$
holds than 
 (\ref{eq: sufficient cond2}) also holds. 

By the assumption, 
$$-\frac{r}{n(t-1)} e^{-\frac{r}{n(t-1)}} = -\beta e^{-\beta} \ge -\frac{1}{e} (1-R)^{\frac{\ln 2}{t-1}}$$
or equivalently 
$$
 \beta e^{-\beta} \le \frac{1}{e} (1-R)^{\frac{\ln 2}{t-1}}
$$
which completes the proof.
\qed

The following two claims are known results related to the Lambert W function and are given as part of the proof of  \autoref{the: lower bound r}. 

\begin{claim}\label{cla: Lambert}\cite[Section IV]{CG96} For any real numbers $ -\frac{1}{e}\le x < 0 $ and~$y$, the equation $ye^y=x$ has exactly two solutions which are given by $y=W_0(x)$ and $y=W_{-1}(x)$, where $W_0$ and $W_{-1}$ are branches of the Lambert W function.
\end{claim}
\begin{claim}\label{cla: Lambert bound}~\cite[Theorem 1]{C13} For any $u>0$ we have that 
 $$-1 -\sqrt{2u} -u < W_{-1} (-e^{-u -1}) <-1 -\sqrt{2u} -\frac{2}{3}u.$$
\end{claim}

\subsection*{Proof of \autoref{the: lower bound r}}
Denote $x = \frac{1}{e} (1-R)^{\frac{\ln 2}{t-1}}$ and $y=\frac{r}{n(t-1)}$, by Claim~\ref{cla: Lambert}, the equation 
 $$-\frac{r}{n(t-1)} e^{-\frac{r}{n(t-1)}} =-ye^{-y}=-x= -\frac{1}{e} (1-R)^{\frac{\ln 2}{t-1}}$$ has exactly two solutions which are $y=-W_0(-x)$ and $y=-W_{-1}(-x)$.
Note that for any $y\ge 1$ the function $-ye^{-y}$ is continuous and monotonically increasing with $y$. Hence, for any given $R=\frac{k}{n}$ only the branch $W_{-1}$ is relevant. This implies that for $r\ge n(t-1)$, Equation (\ref{eq: sufficient cond on r and R}) holds if and only if $y\ge -W_{-1}(-x)$. 

By Claim~\ref{cla: Lambert bound}, we know that $-W_{-1}(-e^{-u-1})< 1+\sqrt{2u}+u$ for any $u>0$. We can rewrite $-x$ as 
$$
 -x=-\frac{1}{e} (1-R)^{\frac{\ln 2}{t-1}} =
 -e^{\frac{\ln 2}{t-1} \ln {(1-R)}-1}
$$ 
with $u = - \frac{\ln 2}{t-1} \ln {(1-R)} > 0$
to obtain 
\begin{align*}
-W_{-1}(-x) & < 1 +\sqrt{2u} + u 
\\ & = 1 + \sqrt{- \frac{2\ln 2}{t-1} \ln {(1-R)}} - \frac{\ln 2}{t-1} \ln {(1-R)}  
\end{align*}
Hence, a sufficient condition for $\E [X^{(r)}] \le n-k$ is that  
$$
y = \frac{r}{n(t-1)}\ge 1 + \sqrt{- \frac{2\ln 2}{t-1} \ln {(1-R)}} - \frac{\ln 2}{t-1} \ln {(1-R)},
$$
or equivalently,
$$
r \ge n(t-1) - n\ln 2\ln(1-R) + n(t-1) \sqrt{- \frac{2\ln 2}{t-1} \ln {(1-R)}}.
$$
\qed

\section{}
\label{appendix_noisy_alpha}
\subsection*{Proof of \autoref{th:noisy_alpha}:}
We denote by $\omega_i$ the probability of collecting an error-free new strand, given that $i-1$ strands were collected. In this case, $\omega_i = \alpha \frac{n-(i-1)}{n} = \alpha\frac{n-i+1}{n}$
We denote by $\omega_i$ the probability of collecting an error-free new strand, given that $i-1$ strands were collected. In this case, $\omega_i=\alpha \frac{n-(i-1)}{n} = \alpha \frac{n-i+1}{n}$. We let $t_i$ be the time to collect a new error-free strand, given $i-1$ such strands were already sampled. Since $t_i$ is geometric random variable it holds that $t_i = \frac{1}{\omega_i}.$
Thus, from the linearity of expectation, we have that 
\begin{align*}
E\left[ \omega_\alpha (n,k) \right] &= \E [t_1 + t_2 +\ldots t_n] - \E [t_k + t_{k+1} +\ldots t_n] 
\\ & = \E [t_1]  + \E[t_2] +\ldots \E[t_n] - \E [t_k] + \E[t_{k+1}] +\ldots \E[t_n] 
\\ & = \frac{n}{\alpha} \left( \frac{1}{1} + \frac{1}{2} + \cdots + \frac{1}{n} \right) - \frac{n}{\alpha} \left( \frac{1}{1} + \frac{1}{2} + \cdots + \frac{1}{n-k} \right)
\\ & = \frac{n}{\alpha} \left( H_n - H_{n-k} \right)
\end{align*}
\qed

\section{}
\label{Appendix: random access example}
\subsection*{Proof of \autoref{th:MDS_random_access}:}

Let $W_{k,n}$ be the random variable that represents the number of samples needed to obtain $k$ distinct coupons where each draw is taken from a pool of $n$ total coupons. We denote by $W_{k,n}(x)$ the generating function for $W_{k,n}$. For $x< \frac{1}{1-\frac{n-k+1}{n}} = \frac{n}{k-1}$ it is known~\cite{PG83} that, 
\begin{align*}
W_{k,n}(x) &= \sum_{r=0}^{x} P [W_{k,n}= r ] \cdot x^r = \prod_{i=1}^{k} \frac{(n-i+1)x}{n-(i-1)x}.
\end{align*} 

Additionally, let $V_{r,n}$ be the random variable that represents the number of distinct coupons in the first $r$ draws, where each coupon is taken from a pool of $n$ total coupons. Note that \begin{align*}
    P[W_{k,n}=r] & = P[V_{r-1,n}=k-1]\cdot P[V_{r,n}=k|V_{r-1,n}=k-1] \\
               & = \frac{n-(k-1)}{n}  P[V_{r-1,n}=k-1],
\end{align*}
and hence, 
\begin{align}\label{eq: V and W relation}
     P[V_{r-1,n}=k-1] = \frac{n}{n-(k-1)} P[W_{k,n}=r].
\end{align}

We let $D_{k,i,n}$ denote the random variable that represents the required number of draws to obtain $k$ distinct coupons or to retrieve coupon $i$ (whichever occurs first), where each draw is taken from a pool of $n$ total coupons. We denote by $D_{k,i,n}(x)$ the generating function for $D_{k,i,n}$.
To this end we define $D_{k,i,n}^{(j)}$ for $0 \le j \le k-1$, to be the random variable that represents the number of samples needed to obtain $j$ distinct coupons (each not equal to the $i$-th coupon), followed by the $i$-th coupon.  
Additionally, let $D_{k,i,n}^{(k)}$, be the random variable that represents the number of samples needed to obtain $k$ distinct coupons (each not equal to the $i$-th coupon). 

It holds that, 
\begin{align*}
P[D_{k,i,n}^{(k)} = r] = \left(1 - \frac{1}{n} \right)^{r} \cdot P [ W_{k,n-1} =r],
\end{align*}
and 
\begin{align*}
D_{k,i,n}^{(k)}(x) &= \sum_{r=0}^{\infty} x^r \cdot P[ D_{k,i,n}^{(k)} = r] 
\\&= \sum_{r=0}^{\infty}  x^r \cdot \left(1 - \frac{1}{n} \right)^{r}  \cdot P [ W_{k,n-1} =r]
\\ & =  W_{k,n-1} \left( \left( 1 - \frac{1}{n}\right) x \right) 
\\ & =  \prod_{\ell=1}^k \frac{(n-\ell)(1-\frac{1}{n})x}{n-1-(\ell-1)(1-\frac{1}{n})x}.
\end{align*}

For $1\le j \le k-1$, using (\ref{eq: V and W relation}), we have that
\begin{align*}
P[D_{k,i,n}^{(j)} = r] & = \left(1 - \frac{1}{n} \right)^{r-1} \cdot \left(\frac{1}{n}\right) \cdot P [V_{r-1,n-1}=j] \\ & = \left(1 - \frac{1}{n} \right)^{r-1} \cdot \left(\frac{1}{n}\right) \cdot \frac{n-1}{n-1-j} \cdot P [ W_{j+1,n-1} =r]
\end{align*}
and,
\begin{align*}
    D_{k,i,n}^{(j)}(x) & = \sum_{r=0}^\infty x^r\cdot P[D_{k,i,n}^{(j)}=r] \\
    & = \sum_{r=0}^\infty x^r\cdot 
     \left(1 - \frac{1}{n} \right)^{r-1} \cdot \left(\frac{1}{n}\right) \cdot \frac{n-1}{n-1-j} \cdot P [ W_{j+1,n-1} =r] \\ 
     & = \frac{n-1}{n-1-j} \cdot \frac{1}{n}\cdot \left(1 - \frac{1}{n} \right)^{-1} \cdot \sum_{r=0}^\infty x^r\cdot \left(1 - \frac{1}{n} \right)^{r} \cdot P [ W_{j+1,n-1} =r] 
     \\ & = \frac{1}{n-1-j} \cdot W_{j+1,n-1} \left( \left( 1 - \frac{1}{n}\right) x \right) \\
     & = \frac{1}{n-1-j}  \prod_{\ell=1}^{j+1} \frac{(n-\ell)\left(1-\frac{1}{n}\right)x}{n-1-(\ell-1)\left(1-\frac{1}{n}\right)x}. 
\end{align*}


Note that, $P[D_{k,i,n}^{(0)}=r] = \frac{1}{n}$ if $r=1$ and otherwise $P[D_{k,i,n}^{(0)}=r] =0$. Therefore, we have that
$D_{k,i,n}^{(0)} (x) = \frac{x}{n}$.
Next, we present $D_{k,i,n}(x)$ as a function of $D_{k,i,n}^{(j)}$ for $0 \le j \le k$. 
\begin{align*}
D_{k,i,n}(x) &= \sum_{r=0}^{\infty} x^r \cdot P [D_{k,i,n} = r ]
\\ & = \sum_{r=0}^{\infty} x^r \sum_{j=0}^{k} P [D_{k,i,n}^{(j)} = r ]
\\ & = \sum_{j=0}^{k} \sum_{r=0}^{\infty}  x^r \cdot P [D_{k,i,n}^{(j)} = r ]
\\ & = \sum_{j=0}^{k} D_{k,i,n}^{(j)} (x).
 \end{align*}

From the above, it can be derived that, 
\begin{align*}
    \E[D_{k,i,n}] & = D_{k,i,n}'(1)  = \frac{1}{n} + \sum_{j=1}^{k-1}\frac{n(n-(j+1))(\psi(-n)-\psi(j+1-n))}{n(n-(j+1))} + \frac{n(n-k)(\psi(-n)-\psi(k-n))}{n} \\
    & = \frac{1}{n} + \sum_{j=1}^{k-1}(\psi(-n)-\psi(j+1-n)) + (n-k)(\psi(-n)-\psi(k-n)),
\end{align*}
where $\psi (z) \triangleq \int_0^\infty \left( \frac{e^{-t}}{t} - \frac{e^{-zt}}{1-e^{-t}}\right) dt$ is the digamma function. In~\cite{H14}, it is claimed that for any $z \in \C$ and $j \in \N $, we have that, 
$\psi (z +j) = \psi(z) + \sum_{h=1}^j \frac{1}{z+h-1},$ which implies that, 
$\psi (-n) - \psi(j-n) =H_{n} -H_{n-j}.$ 

Hence, 
\begin{align*}
\E[\tau_i(\cC)]& = \E[D_{k,i,n}] 
= \frac{1}{n} + \sum_{j=1}^{k-1} \left(H_n -H_{n-j-1}\right)
 + \left(n-k\right) \left(H_n - H_{n-k}\right) 
 \\ & = \frac{1}{n} + (n-1)H_n - (n-k)H_{n-k} -\sum_{j=1}^{k-1}H_{n-j-1}
 \\ & =  \frac{1}{n} + (n-1)H_n - (n-k)H_{n-k} - \left(\sum_{j=1}^{n-2}H_j-\sum_{j=1}^{n-k-1}H_j\right)
 \\ & \stackrel{(a)}{=} \frac{1}{n}+ (n-1)H_n -(n-k)H_{n-k} - \left((n-1)(H_{n-1}-1 - (n-k)(H_{n-k}-1))\right)
 \\ & = \frac{1}{n}+ (n-1)(H_n-H_{n-1}+1) -(n-k)
 \\ & = \frac{1}{n} + \frac{n-1}{n} + (n-1) - n + k
 \\ & = k,
\end{align*}
where $(a)$ follows since for any integer $n>0$ we have that $\sum_{j=1}^n H_j = (n+1)H_n - n$.

\section{}
\label{AppendixC}
\subsection*{Proof of \autoref{th: random access: (2k,k) code}:}

 Let $\cC_k$ be the systematic $(2k,k)$ code that is defined by $\bfU_k=(\bfu_1,\bfu_2,\ldots,\bfu_k)$ and  
    $$\bfX_k = (\bfu_1,\ldots,\bfu_k, \bfu_1+\bfu_2,\ldots, \bfu_{k-1}+\bfu_k, \bfu_k+\bfu_1).$$
    Similarly to \autoref{ex: random access: expect < k=4}, we have that
    \begin{align*}
    P\left[\tau_1(\cC)\ge r\right] & = \sum_{i=1}^{2k-3} P\left[\tau_1(\cC)\ge r | \cE_{r-1}=i\right]\cdot P\left[\cE_{r-1} = i\right],
\end{align*} 
and 
$$
{P[\cE_{r}=i] =} \frac{\binom{2k}{i}}{(2k)^{r}} \sum_{j=0}^{i} \binom{i}{j}(-1)^{i-j} (i-j)^{r}.
$$ 
Since $P\left[\tau_1(\cC)\ge r | \cE_{r-1}=i\right]$ does not depend on $r$, let us denote $P_i = P\left[\tau_1(\cC)\ge r | \cE_{r-1}=i\right]$. We have that  
\begin{align*}
     \E[\tau_1(\cC)] & =  \sum_{r=0}^\infty P\left[\tau_1(\cC)\ge r\right] = 1 + \sum_{r=1}^\infty  \sum_{i=1}^{2k-3}  P_i\cdot P[\cE_{r-1}=i]\\
     & = 1 + \sum_{i=1}^{2k-3} P_i \sum_{r=1}^\infty P[\cE_{r-1}=i] \\
     & = 1 + 
     \sum_{i=1}^{2k-3} P_i \sum_{r=1}^\infty 
     \binom{2k}{i}\sum_{j=0}^i(-1)^{i-j}\binom{i}{j}\left(\frac{j}{2k}\right)^{r-1}
     \\
     & = 1 + \sum_{i=1}^{2k-3}P_i \binom{2k}{i}\sum_{j=0}^{i}(-1)^{i-j}\binom{i}{j}\sum_{r=0}^\infty \left(\frac{j}{2k}\right)^{r} \\& = 
     1 + \sum_{i=1}^{2k-3} P_i \binom{2k}{i}(-1)^i\sum_{j=0}^{i}(-1)^j\binom{i}{j} \cdot \frac{2k}{2k-j} \\
     & = 1 + \sum_{i=1}^{2k-3} P_i\cdot 2k\cdot \binom{2k}{i} \cdot (-1)^i \cdot \frac{(-1)^i}{(2k-i)\binom{2k}{i}}\\ 
      & = 1 + \sum_{i=1}^{2k-3} P_i\cdot   \frac{2k}{(2k-i)}.
\end{align*}
Hence, our goal is to calculate $P_i$. Let $A(2k-1,i)$ be the number of options to draw $r-1$ strands such that $\bfu_1$ cannot be recovered from this set of draws, knowing that the set of different encoded strands that were drawn is of size exactly $i$. Then, we have that $P_i = \frac{A(2k-1,i)}{\binom{2k}{i}}$.

To present a recursive expression for the values $A(2k-1,i)$, we describe an equivalent way to represent the options that contribute to $A(2k-1, i)$. Let $G_k=(V,E)$ be the directed graph with the $2k-1$ nodes that correspond to the symbols in $\cX$ excluding $\bfu_1$. The set $E$ consists of the following edges: 
\begin{itemize}
\item For each $2\le j\le k-1$, the vertex $  \bfu_j+\bfu_{j+1}$ has four outgoing edges. 
Two green outgoing edges to the nodes $\bfu_j$ and $\bfu_{j-1}+\bfu_j$, and two blue outgoing edges to the nodes $\bfu_{j+1}$ and $\bfu_{j+1}+\bfu_{j+2}$ (where $\bfu_{j+2}=\bfu_1$ if $j=k-1$). 
\item There are two blue outgoing edges from $\bfu_1 + \bfu_2$, to the nodes $\bfu_2$ and $\bfu_2+\bfu_3$.
\item There are two green outgoing edges from $\bfu_k + \bfu_0$, to the nodes $\bfu_k$ and $\bfu_
{k-1}+\bfu_k$.
\end{itemize}
\begin{figure}[h!]
    \centering
    \includegraphics[width=0.8\linewidth]{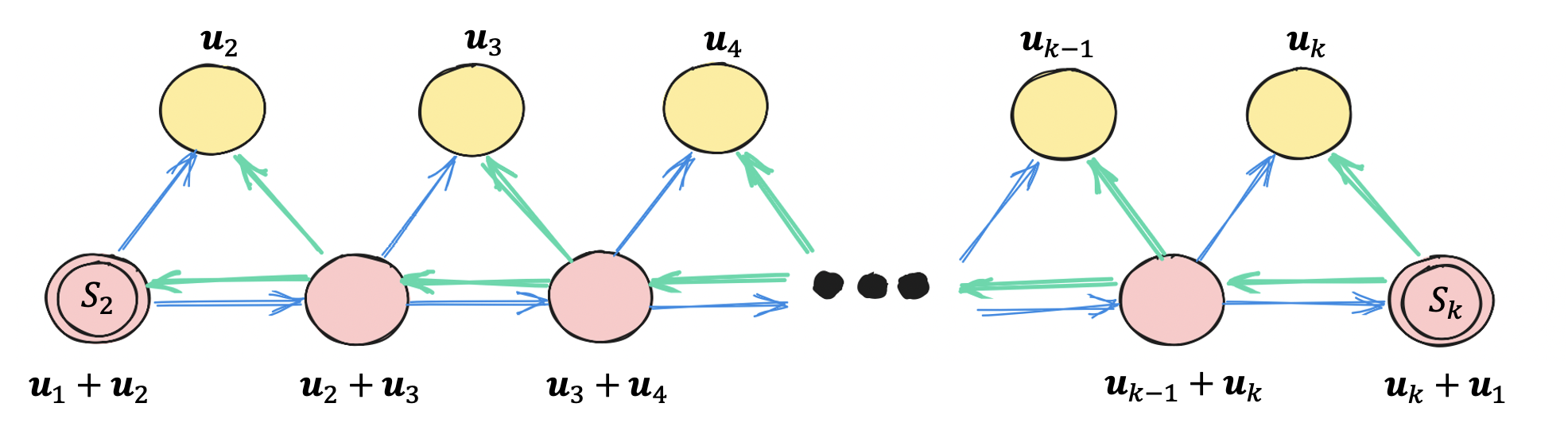}
    \caption{Schematic description of $G_k$}
    \label{fig:graph}
\end{figure}
Denote the nodes $\bfu_1+\bfu_2$ and $\bfu_k+\bfu_1$ by $S_2$ and $S_k$, respectively. Additionally, denote the nodes $\bfu_j$, for $2\le j\le k$ by \emph{ending nodes}. For a set $J\subseteq [2k]\backslash \{1\}$, let $G_k^{(J)}$ be the subgraph of $G_k$ that contains all the nodes that correspond to $J$ (considering their locations in $\bfX$).  
Note that any set $J\subseteq[2k]$ of size $i$ is not a retrieval set of $\bfu_1$ if and only if the subgraph of $G_k^{(J)}$, does not contain a monochromatic path from $S_2$ or $S_{k}$ to one of the ending nodes (if $S_2$, $S_k$ is not in $G_k^{(J)}$, we say that there is no such path from $S_2$, $S_k$, respectively). 
Hence, $A(2k-1,i)$ is equal to the number of subgraphs $G_k^{(J)}$ of $G_k$, such that $J\subseteq [2k]\backslash\{1\}$ and $G_k^{(J)}$ does not contain a monochromatic path from $S_2$ or $S_k$ to one of the ending nodes. Denote the nodes of $G_k^{(J)}$ by $V'$ and consider the following cases.
\begin{enumerate}
    \item If $S_2,S_k\notin V'$ then any such a subgraph $G_k^{(J)}$ cannot contain a valid monochromatic path and there are $\binom{2k-3}{i}$ such subgraphs.
    \item If $S_2\in V'$ then we have that $\bfu_2\notin V'$ and there are $A(2k-3,i-1)$ such sub-graphs.
    \item If $S_k\in V'$ then we have that $\bfu_k\notin V'$ and there are $A(2k-3,i-1)$ such sub-graphs.
    \item If $S_2,S_k\in V'$ then we have that $\bfu_2,\bfu_k\notin V'$ and there are $A(2k-5,i-2)$ such subgraphs. 
\end{enumerate}
Thus, 
\[
A(2k-1,i) = \binom{2k-3}{i} + 2A(2k-3,i-1) - A(2k-5,i-2).
\]
By denoting $B(k,i) = A(2k+1,i)$, we can write the latter as 
\[
B(k,i) =  \binom{2k-1}{i} + 2B(k-1,i-1) - B(k-2,i-2),
\]
for any $k\ge 2, i\ge 2$, and for all $k\ge 0$ we have that $B(k,0)=1$ and $B(k,1)=2k+1$. Additionally $A(1,2)=1$, for $i\ge 2$ we have $A(0,i)=0$ and for $i\ge 3$ we have $B(1,i)=0$.

Thus, we have that 
\[
\E[\tau_1(\cC)] = 1 + \sum_{i=1}^{2k-3} B(k,i) \cdot \frac{2k}{(2k-i)\binom{2k}{i}}.
\]

\end{document}